%% file: main.tex
\documentclass[a4paper,11pt]{article}
\pdfoutput=1 % if your are submitting a pdflatex (i.e. if you have
\usepackage{jheppub} % for details on the use of the package, please
\usepackage[T1]{fontenc} % if needed
\usepackage[utf8]{inputenc}
\usepackage{amsmath,amssymb,amstext,amsfonts} % AMS' mathematical symbols

\usepackage[dvipsnames]{xcolor}
\usepackage{graphicx}
\usepackage{hyperref}
\usepackage{cancel} %for strike-through font
\usepackage{float}
\usepackage{bbm}
\usepackage{mathtools}

\usepackage{lscape}
\usepackage{siunitx}

\makeatletter
\renewcommand*\env@matrix[1][\arraystretch]{%
  \edef\arraystretch{#1}%
  \hskip -\arraycolsep
  \let\@ifnextchar\new@ifnextchar
  \array{*\c@MaxMatrixCols c}}
\makeatother

\makeatletter
\g@addto@macro\bfseries{\boldmath}
\makeatother

\newcommand{\eps}{\epsilon}

\newcommand{\cEf}[4]{{\mathcal{E}_4}\!\left(\begin{smallmatrix}#1\\#2\end{smallmatrix};#3,#4\right)}

\def\beq{\begin{equation}}
\def\eeq{\end{equation}}
\def\bsp#1\esp{\begin{split}#1\end{split}}

\makeatletter
\newcommand{\IEIF}{%
  \def\@IEIFsep{(}%
  I_F\@IEIFi
}
\newcommand\@IEIFi{\@ifnextchar\stopIEIF{\@IEIFend}{\@IEIFii}}
\newcommand\@IEIFii[4]{%
  \big\@IEIFsep
  \begin{smallmatrix}
    #1 & #2 \\
    #3 & #4
  \end{smallmatrix}
  \def\@IEIFsep{|}
  \@IEIFi
}
\newcommand\@IEIFend[2]{% 
  ; #2 \bigr)
}
\makeatother
\newenvironment{gleichung}{\begin{equation}\begin{aligned}}{\end{aligned}\end{equation}\\ \noindent}
\newenvironment{gleichung*}{\begin{equation*}\begin{aligned}}{\end{aligned}\end{equation*}}

 \usepackage{tikz}
    \usetikzlibrary{positioning}
    \usetikzlibrary{arrows}
    \usetikzlibrary{shapes}
    \usepackage{pgfplots}
    \usetikzlibrary{calc}
    \usetikzlibrary{decorations.markings}
     \usetikzlibrary{decorations.pathmorphing}
    \tikzset{snake it/.style={decorate, decoration=snake}}
\def\centerarc[#1](#2)(#3:#4:#5) % Syntax: [draw options] (center) (initial angle:final angle:radius)
    { \draw[#1] ($(#2)+({#5*cos(#3)},{#5*sin(#3)})$) arc (#3:#4:#5); }
    \usepackage{booktabs}

\definecolor{cnblue}{RGB}{7,82,154}

\title{\boldmath 
 On a procedure to derive $\eps$-factorised differential equations 
 beyond polylogarithms}

\author[a]{Lennard Görges} 
\author[a]{Christoph Nega} % \note{Corresponding author.}}
\author[a]{Lorenzo Tancredi}
\author[a]{and Fabian J. Wagner} % \note{Also at Some University.}}

\affiliation[a]{Technical University of Munich, TUM School of Natural Sciences, Physics Department, James-Franck-Straße 1, 85748 Garching, Germany}

\emailAdd{lennard.goerges@tum.de}
\emailAdd{c.nega@tum.de}
\emailAdd{lorenzo.tancredi@tum.de}
\emailAdd{fabianjohannes.wagner@tum.de}

\preprint{TUM-HEP-1458/23}

\abstract{

In this manuscript, we  elaborate on a  procedure to derive $\eps$-factorised
differential equations for multi-scale, multi-loop classes of Feynman integrals
that evaluate to special functions beyond multiple polylogarithms.
We demonstrate the applicability of our approach
to diverse classes of problems, by working out $\eps$-factorised differential equations
for single- and multi-scale problems of increasing complexity. To start we are
reconsidering the well-studied equal-mass two-loop sunrise case, and move then
to study other elliptic two-, three- and four-point problems depending on multiple different
scales.
Finally, we showcase how the same approach allows us to obtain $\eps$-factorised differential 
equations also for Feynman integrals that involve geometries beyond a single elliptic curve.

}

\begin{document} 
\maketitle
\flushbottom

\input{Sec1_Introduction.tex}

\input{Sec2_Procedure.tex}

\input{Sec2.2_Sunrise.tex}

\input{Sec3_Applications.tex}

\input{Sec4_BeyondEll}

\input{Sec5_Conclusions.tex}

\acknowledgments

We thank Claude Duhr, Giulio Gambuti, Albrecht Klemm, Nikolaos Syrrakos, Xing Wang for insightful discussions.
We are especially thankful to Xing Wang for having shared his results on some of the Feynman integrals analyzed in this paper, prior to publication.
This work was supported in part by the Excellence Cluster ORIGINS funded by the
Deutsche Forschungsgemeinschaft (DFG, German Research Foundation) under Germany’s
Excellence Strategy – EXC-2094-390783311 and in part by the European Research Council (ERC) under the European Union’s research and innovation programme grant agreement 949279
(ERC Starting Grant HighPHun).

\appendix

\input{App2_Triangle}

\input{App3_Add}

%---------bibliography if using bibtex
\bibliographystyle{JHEP} % doi-Link, when doi  = {10.1039/C3CC46767H} in .bib file
\bibliography{eps_form} % with XeLaTex: needs the same name as main texfile

\end{document}

%% file: Sec1_Introduction.tex
% !TEX encoding = UTF-8 Unicode

\section{Introduction}
Feynman integrals have played a central role in 
theoretical physics since the early days of application of perturbative Quantum Field Theory to describe natural phenomena.
Together with being indispensable building blocks for the calculation
of many relevant observables both for low- and high-energy physics, their explicit
analytic evaluation has revealed unexpected connections 
with modern topics in pure mathematics such as number theory and algebraic geometry.
This new point of view has in turn helped to clarify how certain general properties of 
scattering amplitudes
can be made manifest at the level of the Feynman integrals used to represent them.
Given the complexity of the calculations required to express scattering amplitudes
in terms of special functions, any insights on the types of mathematical
structures expected to appear can prove crucial to simplify their computation.

Undoubtedly, a very important role in these developments has been played by
the so-called differential equation method~\cite{Kotikov:1990kg,Remiddi:1997ny,Gehrmann:1999as},
which allows one to demonstrate that Feynman integrals satisfy systems of linear differential equations 
with rational coefficients. The most straightforward way to 
derive such differential equations is based on the existence of \emph{integration-by-parts identities}
(IBPs)~\cite{Tkachov:1981wb,Chetyrkin:1981qh} among Feynman integrals. 
IBPs are linear
systems of equations with rational coefficients, which can be solved 
algorithmically~\cite{Laporta:2000dsw} and allow one to express many apparently different
Feynman integrals in terms of so-called \emph{master integrals}. 
Since these master integrals provide in all respects a basis in the vector space 
of Feynman integrals for the problem considered,
an important question becomes whether different choices of bases
can simplify their calculation and highlight the structure of the physical quantities that
one tries to compute.

An important step in this direction was achieved a bit more than a decade ago
with the introduction of the idea of local integrals~\cite{ArkaniHamed:2010gh} as natural
building blocks to represent scattering amplitudes
in the maximally symmetric N=$4$ Super Yang-Mills (SYM) theory. A crucial property of these integrals
is that they can be expressed as iterated integrations over logarithmic differential forms,
with maximum codimension residues normalised to one (or in general to an integer number). These residues
are usually referred to as \emph{leading singularities} and integrals with the property above are said to have
\emph{unit leading singularities}.
Interestingly, it was soon realised that under some conditions, 
these integrals satisfy particularly simple systems of
differential equations which are said to be in \emph{canonical form}~\cite{Henn:2013pwa}.
An important (but not the only) property of such canonical differential equations, is that they
are in \emph{$\eps$-factorised form}, where $\eps$ is the dimensional regularisation
parameter. Since the whole dependence on $\eps$ is factorised in front
of an $\eps$-independent matrix of differential forms, 
this makes the all-order iterative structure of the Feynman integrals as Laurent series in $\eps$ manifest.
The price to pay to achieve this structure is that the differential forms are no longer
guaranteed to be rational functions, but become instead in general transcendental functions. 
It is easy to see that this is the case since the new forms
originate from the solution of the homogeneous 
system of differential equations close to $\eps=0$, which in turn can be 
evaluated using generalised unitarity cuts~\cite{Primo:2016ebd,Bosma:2017ens,Primo:2017ipr}. A particularly powerful technique to compute these cuts involves the use of the Baikov representation~\cite{Baikov:1996iu,Baikov:1996rk,Frellesvig:2017aai,Harley:2017qut}.

As long as we limit ourselves to consider integrals that can be expressed as iterated
integrations over logarithmic differential forms, a lot is understood about how such 
a canonical form can be achieved.
Geometrically, these integrals can often be related to iterated
integrations of rational functions defined on a genus zero surface, i.e. on the Riemann sphere,
and in those cases, they can be expressed in terms of a well-understood class of functions, namely
multiple polylogarithms (MPLs)~\cite{Kummer,Remiddi:1999ew,Goncharov:1998kja,
GoncharovMixedTate,Vollinga:2004sn}.\footnote{Note that this is not always the case, see for example~\cite{Duhr:2020gdd}.}
Various approaches have been developed to achieve a canonical form in these
cases. These include algorithms that aim to transform directly the system of 
differential equations~\cite{Lee:2013hzt,Gituliar:2017vzm,Meyer:2017joq,Dlapa:2020cwj}, 
or that are based on the study of the residues of the corresponding integrands~\cite{Henn:2020lye}.
While none of these techniques is guaranteed to work in all cases, the problem is considered to be 
at least conceptually under control.

The situation is very different once we leave the space of logarithmic differential forms,
increasing the genus and (or) the dimension of the algebraic surfaces involved.
This is an important problem, both from a conceptual and a practical point of view, 
since many examples are known of Feynman integrals which can be naturally defined either on higher-genus two-dimensional
complex surfaces, as elliptic curves, or even on higher-dimensional 
Calabi-Yau geometries, see for example~\cite{Broadhurst:1987ei,Bauberger:1994by,Bauberger:1994hx,Laporta:2004rb,Kniehl:2005bc,Aglietti:2007as,Brown:2013hda,CaronHuot:2012ab,Brown:2010bw,Bloch:2014qca,Bloch:2016izu,Primo:2017ipr,Adams:2018kez,Bezuglov:2021jou,Kreimer:2022fxm,Fischbach:2018yiu,Klemm:2019dbm,Bourjaily:2018yfy,Bourjaily:2019hmc,Bourjaily:2018ycu,Vergu:2020uur,Duhr:2022pch}.

Much progress has been made in the past decade, 
in particular, for what concerns single-scale problems and the genus one case~\cite{Primo:2017ipr,Dlapa:2022wdu,Giroux:2022wav,Pogel:2022yat,Pogel:2022ken,Pogel:2022vat}. Still,
a lot remains to be understood
if either the geometry becomes more complicated or multiple scales are involved.\footnote{For an important exception see~\cite{Adams:2018bsn}.} Moreover, also in those cases where results are known,
there is not yet a consensus on whether the bases identified fulfil the right criteria
to be the natural generalisation of a canonical basis, as originally defined in~\cite{ArkaniHamed:2010gh,Henn:2013pwa}.
From this perspective, one of the most pressing questions
concerns the possibility to develop an approach that allows one to determine bases
of Feynman integrals whose differential equations are in $\eps$-factorised-form and that 
make their iterative
structure manifest, for example avoiding linearly dependent differential forms, 
as much as possible independently of the geometry and of the number of scales involved.

In this paper we try to address this intricate problem, moving from an empirical observation 
made in the context of Feynman integrals which can be evaluated in terms of elliptic Multiple Polylogarithms (eMPLs)~\cite{MR1265553,LevinRacinet,BrownLevin,Broedel:2017kkb}.
In fact, it was observed that whenever explicit results can be obtained in terms of a pure version of eMPLs~\cite{Broedel:2018qkq}, linear combinations of Feynman integrals
can be identified that appear to have the right properties of purity and uniform transcendental
weight expected from canonical Feynman integrals~\cite{Broedel:2019hyg}.\footnote{We stress that a
consensus on a way to define transcendental weight beyond polylogarithms has not yet been reached in the physics and mathematics literature. For practical purposes, here we refer to the way of counting transcendental weight introduced in~\cite{Broedel:2018qkq} and commonly used in the physics literature.} This rotation
involves a specific separation of the period matrix (which solves the corresponding homogeneous differential equations~\cite{Primo:2016ebd}) into a \emph{semi-simple} and a \emph{unipotent} part. 
This  observation has more recently been confirmed in the case of the differential equations of the 
two-loop sunrise graph~\cite{Frellesvig:2021hkr,Frellesvig:2023iwr}. 
The goal of this paper is to show that, 
if one starts from a basis of master integrals that fulfils some minimal, physically motivated requirements, 
this construction turns out to be very general and can be used as the starting point towards a procedure to find $\eps$-factorised bases of master integrals in a large variety of problems, both in the elliptic case and beyond. 

The paper is organised as follows: In section \ref{sec:procedure} we explain our procedure. Our method is based on five main steps which are introduced and discussed. Subsequently, we provide a pedagogical example of our method in section \ref{sec:sunrise}, where we discuss our approach applied to the sunrise family with different mass configurations. We then continue in section \ref{sec:applications} with examples of Feynman graph families with a single underlying elliptic curve. We consider one-, two- and three-parameter families of integrals, including a four-point example at two loops. Afterwards, in section \ref{sec:beyondell} we apply our procedure also to Feynman graph families characterised by more complicated geometries than a single elliptic curve. Finally, we provide our conclusions in section \ref{sec:conclusions}. We also  included several appendices as well as an ancillary file containing explicit results.

%% file: Sec2_Procedure.tex
% !TEX encoding = UTF-8 Unicode

\section{Description of the procedure}
\label{sec:procedure}

We consider dimensionally regularised Feynman integral families of the form 
\begin{equation}
\label{def:integralfamily}
I_{\nu_1,\dots,\nu_{n+m}}(\underline z;d) = \int \left(\prod_{j=1}^l\frac{\mathrm d^dk_j}{i\pi^{d/2}}\right)\frac{\prod_{j=1}^m N_j^{-\nu_{n+j}}}{\prod_{j=1}^n D_j^{\nu_j}}\, ,
\end{equation}
where $\nu_j \leq 0$ for all $j>n$. The set of propagators in the denominator $\{D_1,D_2, \dots, D_n \}$ is specified by the topology of the Feynman graph under consideration. The $\{N_1,N_2, \dots, N_m \}$ are a minimal set of irreducible scalar products in the problem, i.e. scalar products involving loop momenta that cannot be written as a linear combination of the propagators. 
We set $d_0 =2n$ for different 
$n \in \mathbb{N}$, depending on the family under consideration.

As outlined in the introduction, Feynman integrals that belong to a given family, exhibit the structure of a finite-dimensional vector space, whose basis we refer to as
\emph{master integrals}. We indicate them in vector form as $\underline I$. 
Moreover, 
any choice of basis of master integrals $\underline I$ satisfies a system of linear first-order differential equations 
with respect to the kinematic variables and the masses the problem depends on. 
We indicate the set of all variables with $\underline z$.
The system of differential equations is also often referred to as \emph{Gauss-Manin} (GM) differential system and it takes the general form
\begin{equation}
\mathrm d \underline I = \mathbf{GM}(\underline z, \epsilon )    \underline I\,.
\end{equation}
Our goal is to describe a procedure to find a transformation matrix $\mathbf R(\underline z,\eps)$ constructed as a series of subsequent rotations, i.e. 
\begin{gleichung}
\underline J = \mathbf R(\underline z, \eps) \underline I \quad\text{with}\quad \mathbf R(\underline z,\eps) = \mathbf R_r (\underline z,\eps)\cdots \mathbf R_2(\underline z,\eps) \mathbf R_1(\underline z,\eps) \, ,
\end{gleichung}
that cast the system of differential equations into $\eps$-form, namely
\begin{equation}
\mathrm d \underline J = \epsilon\,  \mathbf{GM}^{\eps}(\underline z )    \underline J\, , \quad \mbox{where} \quad 
\epsilon\, \mathbf{GM}^{\eps}(\underline z ) = 
\left[ \mathbf R(\underline z, \epsilon) \mathbf{GM}(\underline z, \epsilon) + \mathrm d \mathbf R(\underline z , \epsilon)  \right] \mathbf R(\underline z, \epsilon) ^{-1}\,. \label{eq:epsform}
\end{equation}
The  crucial property of eq.~\eqref{eq:epsform} is that the new matrix $\mathbf{GM}^{\eps}(\underline z)$ does not depend on $\epsilon$. 
If the Gauss-Manin system is in this particular form we call it to be \emph{$\eps$-factorised}.
In addition, in the polylogarithmic case one can (conjecturally) always bring the Gauss-Manin matrix $\mathbf{GM}^{\eps}(\underline z)$ in an even more specific form, where all its entries $\left[\mathbf{GM}^{\eps}(\underline z)\right]_{ij}$ are given in terms of $\mathrm d\log$-forms. If this is the case, we say that the system is in \emph{canonical form}.

An $\epsilon$-factorised form implies that the iterative structure of the solution is manifest in the differential forms appearing in $\mathbf{GM}^{\eps}(\underline z)$ and, if one
understands all non-trivial relations among these forms, one can claim to fully control the functional relations among the iterated integrals that stem from them. 
This statement can be made more precise as follows: If we call the entries of the matrix 
$\left[\mathbf{GM}^{\eps}(\underline z)\right]_{ij} = \omega_{ij}$, then the condition for the iterated integrals to be linearly independent with respect to some subalgebra of functions $\mathcal{F}$, is that there is no exact form $\eta = \mathrm d f$ with $f \in \mathcal{F}$ such that
\begin{equation}
\sum_{ij} a_{ij} \omega_{ij} = \eta  \quad \mbox{for} \quad a_{ij} \in \mathbb{C}
\label{eq:indepomega}
\end{equation}
with not all $a_{ij}=0$. This condition is then sufficient to guarantee that
the corresponding iterated integrals are linearly independent~\cite{ChenSymbol}.
Clearly, if the forms are all logarithmic, all relations among them descend trivially from corresponding relations among the respective logarithms.

\subsection*{Overview of the procedure}

Before describing the idea in detail, we would like to stress that our procedure is not fully algorithmic: while we have identified
a set of general steps which  work very well 
for all genus one cases we have studied (and also in some cases beyond genus one), 
the implementation of each step requires a case-by-case analysis and their 
complexity depends very much on the problem under consideration. 
We stress that this procedure can be applied to individual sectors of a graph, 
allowing one to work bottom-up, sector-by-sector, to construct 
a basis whose Gauss-Manin system is $\epsilon$-factorised.

Schematically, we can summarise our procedure in five main steps as follows:
\begin{enumerate}
    \item Make a reasonable choice of initial basis for the problem considered.
    \item Calculate the fundamental matrix of solutions for the  homogeneous system of differential equations at $\eps=0$, split it into a \emph{unipotent} and \emph{semi-simple} part, and rotate the initial basis with the inverse of the semi-simple part.
    \item Adjust factors of $\eps$ and swap integrals in the basis such that non $\eps$-factorised terms only appear 
     below the diagonal of the Gauss-Manin connection matrix. We reach in this way an \emph{upper-triangular $\eps$-form}.
    \item Integrate out iteratively the remaining unwanted terms in the homogeneous part of the differential equations. 
    \item Perform shifts in the inhomogeneous part to factorise $\eps$ in the whole system of differential equations.
\end{enumerate}

Let us begin by presenting some general considerations which are behind each of the points above.

\subsubsection*{Step 1: Choice of the initial basis}
The first step is extremely important but it is also the least understood one.
First of all, we should stress the obvious fact that, as long as we assume that a basis of master integrals exists whose differential equations are in $\eps$-factorised form, it is always possible, at least in principle, to reach it starting from any other basis by a sequence of transformations involving arbitrarily complicated functions of $\eps$ and of the kinematics.\footnote{Clearly, as soon as we leave the realm of multiple polylogarithms, a rotation to an $\eps$-factorised basis will necessarily involve transcendental functions.}
As a consequence, there is no unique choice of starting basis for our procedure to work.
However, as in the purely polylogarithmic case, if we start from a basis that is particularly far from a \emph{good basis}, our procedure might either become computationally extremely demanding or even fail at any of the subsequent steps. 
The question becomes therefore, what are general criteria that allow us to say that the basis we are starting from 
is not too far from a  \emph{good basis}.

To this aim, it is useful to start from the much better understood polylogarithmic case. For a rather large class of problems that can be solved in terms of algebraic functions and Chen iterated integrals over $\mathrm d\log$-forms, a study of the integrand can provide important information.\footnote{Note that this goes beyond the realm of multiple polylogarithms, see for example \cite{Duhr:2020gdd}.} In fact, as elucidated in~\cite{ArkaniHamed:2010gh,Henn:2013pwa,Henn:2020lye}, candidates for canonical integrals which fulfil $\eps$-factorised differential equations, can be identified by selecting integrands which can be expressed as iterated $\mathrm d\log$-forms with coefficients equal to  numbers. 
The integrand in this case is said to be in $\mathrm d\log$-form and to have unit leading singularities.
This can be achieved in practice by an analysis of the iterated residues of the corresponding integrands, in any suitable representation (Feynman-Schwinger parameters, Baikov representation, etc). The analysis is usually performed in
$d=d_0 - 2 \eps$ dimensions, with typically $d_0 =2,4,6$. As a matter of fact, studying the integrand exactly in $d=d_0$ provides often enough information to determine a suitable basis for general values of $\eps$.\footnote{One should take extra care if a purely $d_0$-dimensional parametrization of the loop momenta is used, since numerators proportional to Gram determinants would all be identically zero and one would lose possible candidates for canonical integrals. This is avoided using Feynman or Baikov parametrization.} Intuitively, this can be understood
realising that for $d=d_0 - 2 \eps$ the integrand can be schematically parameterised as
\begin{align}
I \sim \int \prod_{i=1}^n \mathrm dx_i\,  \mathcal{F}(x_i, \underline{z} ) \left( \mathcal{G}(x_i, \underline z)\right)^\eps
\end{align}
with $\mathcal{F}(x_i, \underline{z} )$ and $\mathcal{G}(x_i, \underline z)$ algebraic functions, $x_i$ are the integration variables (for example the Baikov parameters) and $\underline z$ the set of kinematical invariants and masses that the integral depends on. Therefore, if the integrand is in $\mathrm d\log$-form 
for $\eps=0$, higher order corrections in $\eps$ will not invalidate this form, 
as they will naturally add only powers of logarithms.

This analysis is extremely powerful and one might wonder how it could be generalised beyond $\mathrm d \log$-forms. In the following sections, we will provide an example of how this could be achieved in the genus one case, building upon the construction of pure elliptic multiple polylogarithms provided in~\cite{Broedel:2018qkq}.
Nevertheless, despite being informative, the study of leading singularities 
 has at least two drawbacks. First of all, in a general multi-parameter case and for increasing numbers of loops, it becomes computationally extremely difficult to analyse all iterated residues.
In these cases, a simpler analysis restricted to some specific subsets of generalised cuts can provide partial but useful information to define good candidates for a canonical integral.
More importantly, depending on the parametrization that one chooses, analysing the residues of the integrand often imposes too restrictive conditions, such that not enough canonical candidates can be identified unless the family of integrals is enlarged. In complicated multi-loop and multi-parameter cases, this is often not practical.

A classical example is provided by integrals with squared propagators.  
Such integrals are often excluded a priori in the residue analysis since squared propagators typically
show up as double poles in the integrand. When dealing with massless Feynman integrals, insisting on the absence
of double propagators is often well justified, since the latter typically generate power-like 
infrared (IR) divergences, which are not expected to appear in gauge theories and which would manifest as 
non-logarithmic singularities in the Feynman integrals.
On the other hand, this analysis is often too superficial, as it does
not take into account the fact that, after integration on the contour that defines the Feynman integral,
these double poles might rearrange into harmless single poles. 
This can be seen already in the case of the one-loop bubble integral family:
\begin{align}
{\rm Bub}_{\nu_1, \nu_2}(p^2,m^2) = \int \frac{\mathrm d^d k}{i \pi^{d/2}} \frac{1}{(k^2 -m^2)^{\nu_1}((k+p)^2-m^2)^{\nu_2}}\,.
\end{align}
It is well known that both for $m^2=0$ or $m^2 \neq 0$, a candidate for a canonical integral in $d=4 -2 \eps$ dimensions
is provided by the integral with a propagator squared, namely ${\rm Bub}_{2,1}(p^2,m^2)$.
The reason why this is a good  candidate in both cases can be seen in various ways, the simplest being that the bubble with a squared propagator in $d=4-2 \eps$ dimensions is proportional to the one with linear propagators in $d=2-2\eps$. 
Importantly, in the massless case, one can always trade the bubble with a squared propagator  with a three-point function, since the two are not independent under integration by parts identities. On the other hand, in the massive case, both master integrals are independent and have to be kept. While this example is extremely simple, it points to the fact that admitting bases of master integrals with squared massive propagators is unavoidable to construct candidates for $\epsilon$-factorised differential equations, even in the genus zero case. 

Moving from these general considerations, we can now summarise our approach to select a good starting point for our procedure.  We  work sector by sector to construct our basis and all integrals are assumed to live in $d=d_0 - 2\eps$ dimensions, with $d_0 =2n$ and $n \in \mathbb{N}$. We notice here that our procedure can be seen as a generalisation of the
one proposed in~\cite{Gehrmann:2014bfa} beyond the polylogarithmic case.

\begin{enumerate}

   \item First and foremost, we always avoid integrals with power-like UV or IR divergences in $d=d_0$.

   \item For sectors with one single master integral, we try to select a candidate with unit leading singularities, at least on the maximal cut. In general, more complicated geometries can be hidden in lower cuts. Since we work bottom-up, we assume that an appropriate basis for them has already been worked out.

   \item For sectors with more than one master, we first focus on the homogeneous part of the  differential equations in strictly $d=d_0$. Either by studying the maximal cut of the master integrals in the sector or, equivalently, by analysing the Picard-Fuchs operator associated with the homogeneous system, we determine the geometry associated with that sector (genus zero, genus one, Calabi-Yau etc.). This informs us on the minimal irreducible complexity in the corresponding block of differential equations, i.e., how many master integrals can be decoupled by just rational or algebraic transformations.

    \item If the analysis at point 3. reveals that the sector can be completely decoupled, we attempt to choose integrals with unit leading singularities, following the standard approach.

    \item If the analysis at point 3. reveals instead that $n$ integrals are coupled even in $d=d_0$ and if the sector contains $m > n$ integrals in $d=d_0-2 \eps$, we first choose the remaining $m-n$ integrals to make this minimally coupled system manifest. This can be often achieved by selecting integrals which are zero in $d=d_0$, by introducing
    numerators build from Gram determinants of loop and external momenta, see for example~\cite{Remiddi:2013joa, Tancredi:2015pta, Duhr:2022dxb}.  In addition, we also select as many of the extra master integrals as possible, such that their maximal cut can be still localised by taking residues. In the elliptic case, this corresponds to selecting differentials of the third kind.

    \item After decoupling, 
    we are left with choosing a basis for the minimally coupled sectors. In this case, we proceed as follows: First (if necessary, restricting the analysis to the maximal cut) we always choose the first integral to be expressed as a series of $\mathrm d \log$-forms up to at most one integration. For the latter, we impose that it corresponds to the holomorphic differential of the first kind on the geometry considered. 
    In the elliptic case, integrating this form on the two independent cycles provides the two periods of the elliptic curve. Similarly, on more complicated geometries, one obtains one set of independent homogeneous solutions for the corresponding differential equations. One reasoning behind choosing one integral in this form is that integrating over the holomorphic differential form does not introduce any power-like IR divergences. Moreover, as it was worked out explicitly in the elliptic case, this form is one of the integration kernels that we always expect in generalisations of polylogarithms beyond genus zero~\cite{Broedel:2018qkq}. Notice that this differential form does not need to show up in the last integration.
    
    \item The last problem is how to choose the remaining integrals in the coupled sectors, since only excluding power-like IR or UV behaviour usually does not completely constrain the basis choice. In the elliptic case, one possibility could be to choose as extra candidates, integrands which can be written as strings of $\mathrm d \log$s and $\partial\cEf{m}{n}{x}{\vec{a}}/\partial x$, i.e.,
    kernels that define pure elliptic polylogarithms, see~\cite{Broedel:2018qkq}. 
    This can be attempted in simple examples, 
    but the analysis required quickly becomes cumbersome for multi-loop, multi-scale problems.
    Therefore, we choose the extra masters as linear combinations of iterated derivatives of the first integral with respect to the internal masses. This has the advantage of producing differential equations in standard form, without increasing the degree of divergence in the IR. 

   \item Finally, as a rule of thumb based on experience, we always avoid bases of integrals that generate differential equations whose coefficients have a non-trivial dependence on $\eps$ in the denominators.\footnote{Here we mean poles in $\eps$ or denominators of the form $1/f(\underline z,\eps)$ for some polynomial $f$.}  
   Most of the time, imposing the criteria at points $1.$ to $7.$ already guarantees that this requirement is satisfied, modulo trivial rescaling in $\eps$ of the basis elements.
   If that is not the case, and we still have freedom in choosing our basis, we do that by trial and error avoiding such poles as they increase the complexity of the next steps drastically. 
\end{enumerate}

In conclusion, the points above provide us with a guide to constructing a starting basis of master integrals which does not have undesirable properties, among which the most important ones are power-like divergences and  non-minimal couplings in the homogeneous blocks of the individual sectors, in the limit $d = d_0$. We do not claim that all these points are new elements introduced in this paper but are instead the result of the experience collected by many research groups working on this topic.

\subsubsection*{Step 2: Rotation by the inverse of the semi-simple part of the period matrix}

The second step is the crucial one in our procedure. For a choice of basis with no poles in $\eps$ in the Gauss-Manin connection matrices, we start by computing the fundamental matrix of solutions $\textbf{W}$ at $\eps=0$ for every coupled block in the sector under consideration, discarding any contributions from integrals whose own differential equations do not couple to the block. In the following, we refer to this system also as the \emph{maximal cut system}, as the maximal cuts of the integrals provide a solution to it~\cite{Primo:2016ebd,Primo:2017ipr,Bosma:2017ens,Frellesvig:2017aai,Harley:2017qut}, and to $\textbf{W}$ also as the \emph{Wronskian} matrix or \emph{period} matrix. Beyond the polylogarithmic case, $\textbf{W}$ contains new classes of transcendental functions for which we can always derive a representation in terms of locally convergent power series, containing also logarithmic contributions in the parameters. Furthermore, we note that in the construction of $\textbf{W}$, it is convenient to order the solutions such that the powers of logarithms appearing in their power series expansions increase from left to right in the first row. This is usually referred to as a \emph{Frobenius basis}.

To proceed, we take inspiration from the patterns observed in~\cite{Broedel:2018qkq,Broedel:2019hyg}, where \emph{pure} and \emph{uniform transcendental weight} elliptic Feynman integrals could be selected by a rotation of the basis of integrals
by the so-called \emph{semi-simple} part of the period matrix.
In those references, the observation was based on explicit results obtained by direct integration and therefore was limited to at most two orders in $\eps$.
Our goal here is to generalise this procedure to all orders in $\eps$,  working at the level of the differential equations. We then split $\textbf{W}$ into a \emph{semi-simple} part $\textbf{W}^{\text{ss}}$ and \emph{unipotent} part $\textbf{W}^{\text{u}}$, i.e.
\begin{equation}
    \textbf{W} = \textbf{W}^{\text{ss}} \cdot \textbf{W}^{\text{u}} \, .
\end{equation}
In general, this splitting is not unique. The only requirements are that the unipotent part $\textbf{W}^{\text{u}}$ has to satisfy a unipotent differential equation~\cite{Brown:coaction}, i.e. of the type
\begin{equation}
\mathrm{d} \textbf{W}^{\text{u}} =
\left(
\sum_i \textbf{U}_i (\underline z) \ \mathrm{d} z_i
\right)
\textbf{W}^{\text{u}},
\end{equation}
where the matrices $ \textbf{U}_i (\underline z) $ are nilpotent matrices, while  the semi-simple matrix should only be invertible. For the purpose of our procedure, we perform the splitting in such a way that the semi-simple part $\textbf{W}^{\text{ss}}$ has lower triangular form, while the unipotent piece $\textbf{W}^{\text{u}}$ has upper triangular form, with its diagonal containing only constant entries, normalised to one for convenience. We expect such a  splitting to always be possible for a choice of initial basis following the guidelines outlined in the previous paragraph. We then rotate the basis in the top sector with the inverse of $\textbf{W}^{\text{ss}}$, discarding the contributions of the solutions contained in the unipotent part.

\subsubsection*{Step 3: Transformation to upper triangular $\eps$-form}

The third step of the procedure involves typically nothing more than just adjusting some factors of $\eps$, as well as potentially swapping the position of some of the integrals in the basis, such that terms that are not $\eps$-factorised appear only below the diagonal of the new Gauss-Manin connection matrix $\mathbf{GM}$. Further, all of these terms are either constant or go with inverse powers of $\eps$. Notice that at this point, poles in $\eps$ typically appear. 

\subsubsection*{Step 4: Clean up of the homogeneous differential equations}

As the next step, we integrate out these remaining terms systematically in each homogeneous block. This is achieved by shifting the master integrals which generate non $\eps$-factorised homogeneous equations, by terms proportional to the other master integrals in the same sector. Frequently, this amounts to removing total derivatives involving the objects already introduced by the rotation by the inverse of $\textbf{W}^{\text{ss}}$. However, for more complex problems it happens that not all non-$\eps$-factorised terms can be removed in this way. In these cases, it becomes necessary to introduce new functions, which are defined as (iterated) integrals of products of rational or algebraic functions and the transcendental functions introduced by $\textbf{W}^{\text{ss}}$. This leads to new types of integration kernels in the final $\eps$-factorised differential equations.  It is desirable that these new objects are linearly independent under integration by parts identities as discussed around eq.~\eqref{eq:indepomega}. We will comment on this below when looking at explicit cases.

\subsubsection*{Step 5: Clean up of the inhomogeneous differential equations}
After having completed the $\eps$-factorisation of all homogeneous blocks, the factorisation of the whole system can be achieved by performing suitable shifts in the definitions of the master integrals of a given sector, by  integrals in lower sectors. This is in principle a straightforward procedure. In practice, however, it can become tricky as also the functional dependence on $\eps$ of the coefficient functions in these shifts might be complicated. Moreover, if a given subsector exhibits singularities which were not present in the homogeneous equations for the sector considered, the introduction of new functions might be required. These are likewise (iterated) integrals of functions already present in the differential equations.

%% file: Sec2.2_Sunrise.tex
\section{Application to the two-loop sunrise family}
\label{sec:sunrise}

%-----Picture sunrise Graph-------------
\begin{figure}[!h]
\centering
\begin{tikzpicture}
\coordinate (llinks) at (-2.5,0);
\coordinate (rrechts) at (2.5,0);
\coordinate (links) at (-1.5,0);
\coordinate (rechts) at (1.5,0);
\begin{scope}[very thick,decoration={
    markings,
    mark=at position 0.5 with {\arrow{>}}}
    ] 
\draw [-, thick,postaction={decorate}] (links) to [bend right=0]  (rechts);
\draw [-, thick,postaction={decorate}] (links) to [bend left=85]  (rechts);
\draw [-, thick,postaction={decorate}] (llinks) to [bend right=0]  (links);
\draw [-, thick,postaction={decorate}] (rechts) to [bend right=0]  (rrechts);
\end{scope}
\begin{scope}[very thick,decoration={
    markings,
    mark=at position 0.5 with {\arrow{<}}}
    ]
\draw [-, thick,postaction={decorate}] (links) to  [bend right=85] (rechts);
\end{scope}
\node (d1) at (0,1.1) [font=\scriptsize, text width=.2 cm]{$k_1$};
\node (d2) at (0,0.25) [font=\scriptsize, text width=.2 cm]{$k_2$};
\node (d3) at (0.75,-0.6) [font=\scriptsize, text width=3 cm]{$k_1+k_2-p$};
\node (p1) at (-2.0,.25) [font=\scriptsize, text width=1 cm]{$p$};
\node (p2) at (2.4,.25) [font=\scriptsize, text width=1 cm]{$p$};
\end{tikzpicture}
\caption{The two-loop sunrise graph.}
\label{fig:sunrise}
\end{figure}
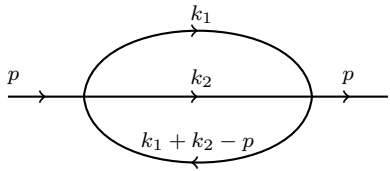
%------------------------

\noindent
Let us see explicitly how our procedure works in the case of the two-loop sunrise 
graph~\cite{Bauberger:1994by,Bauberger:1994hx,Bauberger:1994nk,Caffo:1998du,Laporta:2004rb,Kniehl:2005bc,
Remiddi:2016gno,Adams:2013nia,Adams:2014vja,Adams:2015gva,Adams:2016xah,Bloch:2013tra} shown in figure \ref{fig:sunrise}. 
We work as it is customary in $d=2-2\eps$ 
space-time dimensions and, as it is well known, the result in $d = 4 -2 \eps$ can be obtained by 
dimensional shift~\cite{Tarasov:1996br}.  
We take the associated integral family in the general case of different masses to be defined 
by the following three propagators and two irreducible numerators:
\begin{gleichung}
\label{def_sunrisefamily}
D_1 &= k_1^2-m_1^2\,, \quad &D_2 &= k_2^2-m_2^2\,, \quad D_3 &= (k_1+k_2-p)^2-m_3^2\,, \\
N_1 &= (k_1-p)^2\,, \quad &N_2 &= (k_2-p)^2 \, .
\end{gleichung}
We assume $m_1,m_2,m_3 >0$ and, in the remainder of this section, we set $p^2=s$.

As the first step, we analyse 
the integrand of the so-called \emph{corner} integral $I_{1,1,1,0,0}$ to determine the complexity of the problem.
We use a loop-by-loop Baikov representation~\cite{Baikov:1996iu,Baikov:1996rk,Frellesvig:2017aai,Harley:2017qut} 
in two space-time dimensions. Working out the change of variables we find
\begin{equation}
\label{eq:sunriseBaikov}
    I_{1,1,1,0,0} \sim \int \mathrm{d}z_1 \, \mathrm{d}z_2 \, \mathrm{d}z_3 \, \mathrm{d}z_4 \ 
    \frac{[\mathcal{B}(z_1,z_2,z_3,z_4)]^{-1/2}}{z_1 z_2 z_3}
\end{equation}
with
\begin{equation}
\begin{aligned}
    \mathcal{B}(z_1,z_2,z_3,z_4) = & \left(\left(-s+m_1^2+z_1\right)^2-2 \left(s+m_1^2+z_1\right) z_4+z_4^2\right)  \\ 
    & \quad \times \left(\left(m_2^2-m_3^2+z_2-z_3\right)^2-2
   \left(m_2^2+m_3^2+z_2+z_3\right) z_4+z_4^2\right).
\end{aligned} \label{eq:baikovsun}
\end{equation}
From this representation, it is easy to see that the maximal cut is proportional to a one-fold integral over the reciprocal of the square-root of a quartic polynomial  in the integration variable $z_4$
\begin{equation}
\label{eq:SunriseMaxCut}
    \text{MaxCut}(I_{1,1,1,0,0}) \sim \int \frac{\mathrm{d}z_4}{\sqrt{P_4(z_4)}} \quad\text{with}
\end{equation}
\begin{equation}
\label{eq:quarticpolsun}
    P_4(z_4) = (z_4-a_1) (z_4-a_2) (z_4-a_3) (z_4-a_4) \, ,
\end{equation}
\begin{equation}
\label{eq:rootsun}
    a_1 = \left(m_2 - m_3\right)^2 \, , \ a_2 = \left(m_2+m_3\right)^2\, , \ a_3 = \left(m_1-\sqrt{p^2}\right)^2\, , \ a_4 = \left(m_1+\sqrt{p^2}\right)^2\, . 
\end{equation}
As long as none of the internal masses vanishes, all four roots are distinct and  the geometry is that of an elliptic curve. If at least one of the internal masses is zero, two of the roots degenerate and the elliptic curve reduces to a genus zero surface. 
In the following, we consider three cases explicitly:
\begin{enumerate}
    \item Two equal non-vanishing masses $0 \neq m_2^2=m_3^2\eqqcolon m^2$ and one vanishing mass $m_1^2=0$. This case is polylogarithmic and we can see how our approach works in the case of genus zero geometries.
    \item Three equal non-vanishing masses $0 \neq m_1^2=m_2^2=m_3^2\eqqcolon m^2$. This is the simplest genuine elliptic case.
    \item Three non-vanishing masses of which two are equal, $0 \neq m_1^2 \neq m_2^2=m_3^2  \neq 0$. This case is multivariate and of similar complexity as the generic case of three different internal masses but with considerably more compact mathematical expressions.
\end{enumerate} 

\subsection{The sunrise with two equal non-vanishing masses and a vanishing mass}
For $m_2^2=m_3^2\eqqcolon m^2$ and $m_1^2=0$, the sunrise integral family admits three master integrals, a tadpole and two in the top sector. In this case, the underlying geometry is a Riemann sphere, so if we were to follow our recipe above, we should attempt to identify integrals of unit leading singularity in $d=2$ space-time dimensions. It is immediate to do this analysis directly or using for example \textsc{DlogBasis}~\cite{Henn:2020lye}. Three candidates are quickly found to be
\begin{equation}
\label{masters_eqsunPL}
M_1 = \eps^2 \ I_{1,1,0,0,0}\, , \quad M_2 = \eps^2 \, r(s,m^2)  \, I_{1,1,1,0,0}\, , \quad M_3 =  \eps^2 \left( I_{1,1,1,-1,0} - s\, I_{1,1,1,0,0} \right)\, ,
\end{equation}
involving the square-root 
\begin{equation}
\label{eq:sqrtsun}
        r(s,m^2) = \sqrt{s(s-4m^2)} \, .
\end{equation}
associated to the threshold for the production of two massive and one massless particle. We chose it such that it is manifestly real above the threshold $s\geq 4m^2$. The factor $\eps^2$ is purely conventional such that the expansions of all integrals start at order $\eps^0$. It is easily checked that these master integrals satisfy indeed differential equations in canonical form.

Instead of following this standard approach, let us pretend that we were unable to identify all these candidates and their normalisation factors in the above analysis and use this example to illustrate the steps involved in our procedure. As we will see, they yield essentially the same canonical basis. We start from the basis of master integrals $\underline I=\{I_1,I_2,I_3\}$ given by
\begin{equation}
I_1 = I_{0,1,1,0,0}\, , \quad I_2 = I_{1,1,1,0,0} \quad\text{and}\quad I_3 = I_{1,1,2,0,0} \, .
\end{equation}
They were chosen based only on the following knowledge of their properties:
\begin{enumerate}
\item The tadpole integral $I_1$ is trivial and known to evaluate to a pure function of uniform transcendental weight in $d=d-2 \eps$. 
\item The first integral in the coupled sector, $I_2$, has a representation in terms of iterated $\mathrm d \log$ forms in $d=2$ space-time dimensions.
\item The second master, $I_3$, is proportional to the derivative of the first one with respect to the internal mass squared. In practice, this means that we put a dot on one of its two massive propagators. This guarantees that no spurious power-like IR divergences are introduced.
\end{enumerate}
 The system of non-$\eps$-factorised differential equations for our starting basis $\underline I$ reads
\begin{gleichung}
\label{eq:DEsunrisePLstart}
     \mathrm d \underline I &= \left(\mathbf{GM}_{m^2}\, \mathrm d m^2 + \mathbf{GM}_{s}\, \mathrm d s \right) \underline I \, , \\
\mathbf{GM}_{m^2} &= 
\left(
\begin{array}{ccc}
 -\frac{2 \eps }{m^2} & 0 & 0 \\
 0 & 0 & 2 \\
 \frac{\eps ^2}{m^4 (s-4 m^2)} & \frac{(1+2\eps) (1+3\eps)}{m^2 (s-4 m^2)} & -\frac{s-10 m^2 +\eps  (s-16 m^2)}{m^2 (s-4 m^2)} \\
\end{array}
\right) \, .
\end{gleichung}
For simplicity, we work with the Gauss-Manin connection matrix with respect to $m^2$. The one with respect to $s$ follows at each step from a scaling relation implied by dimensional analysis and Euler's theorem on homogeneous functions\footnote{Alternatively, one might also reduce this to a one-variable problem by working with the ratio of the two scales, $s/m^2$.}. Concretely, we have in~\eqref{eq:DEsunrisePLstart} 
\begin{equation}
    m^2 \, \mathbf{GM}_{m^2} + s \, \mathbf{GM}_{s} =
    \begin{pmatrix}
        -2\eps & 0 & 0 \\
        0 & -1-2\eps & 0 \\
        0 & 0 & -2-2\eps
    \end{pmatrix}.
\end{equation}

The next step in our procedure is to analyse the maximal cut system, which corresponds to the following set of differential equations
\begin{equation}
\label{sunPLinit}
\frac{\partial}{\partial m^2} 
\left(
\begin{array}{cc}
\mathfrak I_2 \\
\mathfrak I_3 \\
\end{array}
\right)
= 
\left(
\begin{array}{cc}
 0 & 2 \\
 \frac{1}{m^2(s-4 m^2)} & \frac{-s+10 m^2}{m^2(s-4m^2)} \\
\end{array}
\right)
\left(
\begin{array}{cc}
\mathfrak I_2 \\
\mathfrak I_3 \\
\end{array}
\right).
\end{equation}
Its fundamental matrix of solutions $\textbf{W}$ can be chosen to take the following form\footnote{If we had started with a better choice of initial master integrals $\underline I$, i.e. $I_{1,1,1,-1,0}$ instead of $I_{1,1,2,0,0}$, the Wronskian matrix \eqref{Wex1} would have been simpler. Here, we choose a non-optimal basis intentionally to illustrate more features of the procedure.}
\begin{align}
\textbf{W} &= \begin{pmatrix}
                     1 & 0 \\
                     0 & \frac{1}{2}
             \end{pmatrix} 
             \begin{pmatrix}
                     \varpi_0 & \varpi_1 \\
                     \partial_{m^2} \varpi_0 & \partial_{m^2} \varpi_1
             \end{pmatrix} \quad\text{with} \nonumber \\
    \varpi_0(s,m^2) &= \frac{1}{r(s,m^2)} =\frac{1}{s} \left[1+2\, \frac{m^2}{s}+6 \left(\frac{m^2}{s}\right)^2+20 \left(\frac{m^2}{s}\right)^3 + \mathcal O\left(\left(\frac{m^2}{s}\right)^4\right)\right] \, , \nonumber \\
    \varpi_1(s,m^2) &= \frac{1}{r(s,m^2)}\log \left(\frac{s-r(s,m^2)}{s+r(s,m^2)}\right) =\varpi_0(s,m^2)\log\left(\frac{m^2}{s}\right) \label{Wex1} \\
    & \quad  +\frac{1}{s}\left[2 \, \frac{m^2}{s}+7 \left(\frac{m^2}{s}\right)^2+\frac{74}{3} \left(\frac{m^2}{s}\right)^3 + \mathcal O\left(\left(\frac{m^2}{s}\right)^4\right) \right]\, , \nonumber
\end{align}
where $r(s,m^2)$ was defined in eq.~\eqref{eq:sqrtsun}. 
In principle, it is possible to rotate the basis of the top sector with the inverse of $\textbf{W}$ to solve the system from eq. \eqref{sunPLinit} at $\eps=0$. However, this step does not lead to a canonical basis, not even to a factorisation of $\eps$. Moreover, the coefficient functions in the differential equations will be of mixed transcendental weight. Instead, we follow \textbf{Step 2} of our procedure and perform the splitting of $\textbf{W}$ into a lower-triangular semi-simple part $\textbf{W}^{\text{ss}}$ and an upper-triangular unipotent part $\textbf{W}^{\text{u}}$. This works because our choice for the first integral in the top sector $I_2$ already has uniform transcendental weight as we anticipated from the properties of its integrand.  Explicitly, we find
\begin{equation}
\textbf{W}^{\text{ss}} = 
\left(
\begin{array}{cc}
 \frac{1}{r(s,m^2)} & 0 \\
 \frac{s}{r(s,m^2)^{3}} & \frac{1}{2m^2 (s-4 m^2)} \\
\end{array}
\right)
\quad \text{and}\quad
\textbf{W}^{\text{u}} =
\left(
\begin{array}{cc}
 1 & \log \left(\frac{s-r(s,m^2)}{s+r(s,m^2)}\right) \\
 0 & 1 \\
\end{array}
\right)
\end{equation}
such that $\textbf{W}^{\text{u}}$ satisfies the unipotent differential equation 
\begin{equation}
\mathrm d \textbf{W}^{\text{u}} =
\left(
\begin{array}{cc}
 0 & \mathrm{d} \log \left(\frac{s-r(s,m^2)}{s+r(s,m^2)}\right) \\
 0 & 0 \\
\end{array}
\right)
\textbf{W}^{\text{u}}\, .
\end{equation}
By splitting the Wronskian matrix $\textbf{W}$ in this way, the logarithm appears exclusively in the unipotent part, while the semi-simple part has uniform transcendental weight zero. Rotating now the basis in the topsector only with the inverse of $\textbf{W}^{\text{ss}}$, we arrive at a new basis
\begin{equation}
    \underline I' =
    \begin{pmatrix}
        1 & \begin{matrix} 0 \ & \ 0 \end{matrix} \\
        \begin{matrix} 0 \\ 0 \end{matrix} & (\textbf{W}^{\text{ss}})^{-1}
    \end{pmatrix}
    \underline I \, ,
\end{equation}
which satisfies the differential equation
\begin{equation}
    \frac{\partial}{\partial m^2} \underline I' =
    \left(
    \begin{array}{ccc}
    -\frac{2 \eps }{m^2} & 0 & 0 \\
    0 & 0 & \frac{r(s,m^2)}{m^2 (s-4m^2)} \\
    \frac{2 \eps ^2}{m^2} & \frac{8\eps (s-m^2) }{(s-4m^2)\, r(s,m^2)}+\frac{12 \eps ^2}{r(s,m^2)} & \frac{\eps (16 m^2 - s)}{m^2 (s-4 m^2)} \\
    \end{array}
    \right)    
    \underline I' .
\end{equation}

The remaining steps to factor out $\eps$ in this problem are now straightforward. First, we rescale the last integral in the basis by a factor $1/ \eps$ with respect to the other two. An additional factor $\eps^2$ is added to all integrals for the same conventional reasons as for the basis in eq.~\eqref{masters_eqsunPL}. This leaves us with just a single term that is not yet proportional to $\eps$. It is, however, a total derivative of an algebraic function already appearing in the problem and it is therefore easily integrated out. These manipulations can be summarised in the rotations
\begin{equation}
   \mathbf{T} =  
          \begin{pmatrix} 
          1 & 0 & 0\\ 0 & 1 & 0 \\ 0 & -\frac{2 (s+2 m^2)}{r(s,m^2)} & 1
          \end{pmatrix}
          \begin{pmatrix} 
          \eps^2 & 0 & 0\\ 0 & \eps^2 & 0 \\ 0 & 0 & \eps 
          \end{pmatrix}\, ,
\end{equation}
which allows us to define a basis that satisfies canonical differential equations
\begin{align}
    \mathrm{d} \underline J &= \eps \ \textbf{GM}^{\eps} \underline J \quad\text{with}\quad \underline J = \left(J_1,J_2,J_3\right)^T = \mathbf{T} \, \underline I' \, , \nonumber \\ \label{def:cbasissunPLus}
    \textbf{GM}^{\eps} &=
    \left(
    \begin{array}{ccc}
    -2 \alpha_1  & 0 & 0 \\
    0 & 2 \alpha_1 - \alpha_2 -3  \alpha_3 \  & \alpha_4 \\
    2 \alpha_1 - 2 \alpha_2 & -6 \alpha_4 & - 3 \alpha_1 + \alpha_2 \\
    \end{array}
    \right) \, ,  \\
    \alpha_1 = \mathrm{d}\log( m^2 )\, , \ \ &\alpha_2 = \mathrm{d}\log (s)\, , \ \ \alpha_3 = \mathrm{d}\log \left(s-4m^2\right)\, , \ \ \alpha_4 = \mathrm{d}\log \left( \frac{s-r(s,m^2)}{s+r(s,m^2)} \right)\, . \nonumber
\end{align}
As advertised, the relation to the canonical master integrals found by \textsc{DlogBasis} is a simple constant rotation that is given by
\begin{equation}
J_1 = M_1 \, , \quad J_2 = M_2\, , \quad J_3 = - M_1 + 3 M_3 \, .
\end{equation}
We would like to stress the following points concerning the application of our procedure in the polylogarithmic case:
\begin{itemize}
    \item The critical step was the splitting of the Wronskian in the semi-simple and unipotent part. Discarding the latter, allowed us to remove the logarithm appearing in the Wronksian matrix, which is the function that would have ruined the transcendental weight properties and the iterative structure of the differential equations.
    \item In general, we do not expect our procedure to outperform other approaches to find canonical bases in the polylogarithmic case. In particular, for an increase in the number of master integrals, the steps involved also increase in complexity. However, if analysing the leading singularities does not provide enough candidates, 
    our method could be used complementary.
\end{itemize}

\subsection{The sunrise with three equal non-vanishing masses}
We now take one step up in complexity to the simplest elliptic scenario and set all three internal masses of the sunrise to the same value, i.e. $m_1^2=m_2^2=m_3^2\eqqcolon m^2$. The number and sector distribution of master integrals does not change compared to the previous case. 

First, we need to choose a basis of integrals to start from. Taking the limit for equal masses in eq.~\eqref{eq:baikovsun}, it is 
easy to study all iterated residues of the integrand corresponding to $I_{1,1,1,0,0}$. 
The complexity of this analysis depends in general on the order in which the various residues are taken. For this particular problem, it is convenient to take residues in the natural order $z_1$ to $z_4$. For the first three integration variables, one easily sees that the integrand has only simple poles in the corresponding variable, such that as expected the result can be written schematically as follows:
\begin{equation}
I_{1,1,1,0,0} \sim \int    \frac{\mathrm{d}z_4}{\sqrt{P_4(z_4)}} \wedge \mathrm{d}\log{f_3(z_3,z_4)} \wedge \mathrm{d}\log{f_2(z_2,z_3,z_4)} \wedge \mathrm{d}\log{f_1(z_1,z_2,z_3,z_4)}\,  , \label{eq:sunLS}
\end{equation}
where the $f_i$ are algebraic functions, at most quadratic in the integration variables $z_1,z_2$ and $z_3$.
If the last integration in $z_4$ were missing, this
integral would be a series of iterated integrations over 
$\mathrm d\log$ forms and would therefore have the right characteristics to be
a candidate canonical integral in the language introduced
in~\cite{ArkaniHamed:2010gh}.

Interestingly, as already hinted at in the previous section, eq.~\eqref{eq:sunLS}
shows that the last integration involves only the holomorphic differential form of 
the first kind on the elliptic
curve, which corresponds exactly to one of the integration kernels that define 
elliptic multiple polylogarithms, as it is manifest in the form originally proposed in~\cite{Broedel:2018qkq}. 
Elliptic multiple polylogarithms can in fact be defined as iterated integrations of rational functions on an elliptic curve as follows:
\beq\label{eq:cE4_def}
\cEf{n_1 & \ldots & n_k}{c_1 & \ldots& c_k}{x}{\vec{a}} = \int_0^x \mathrm du\,\Psi_{n_1}(c_1,u,\vec a)\,\cEf{n_2 & \ldots & n_k}{c_2 & \ldots& c_k}{u}{\vec a}
\eeq
with $n_i\in\mathbb{Z}$, $c_i\in\widehat{\mathbb{C}}$ and $\vec{a} = \{a_1,...,a_4\}$ is the vector of roots defined in eq.~\eqref{eq:rootsun}.
The first integration kernel for $n=0$ reads explicitly
\beq\label{eq:pure_psi0}
\Psi_0(0,x,\vec a) = \frac{c_4}{\varpi_0\,y} = \frac{c_4}{\varpi_0} \frac{1}{\sqrt{P_4(z_4)}}\,,
\eeq
where $y = \sqrt{P_4(z_4)}$ defines the elliptic curve, 
$\varpi_0$ is the holomorphic period on the curve and $c_4$ is
an algebraic function which is added for convenience.
With these definitions, and assuming for the sake of the argument 
that the branching points of the elliptic curve $\vec{a}$ are constant,  
we can schematically write
\begin{equation}
I_{1,1,1,0,0} \sim \varpi_0\int    \mathrm{d} \cEf{0 }{0}{z_4}{\vec{a}} \wedge \mathrm{d}\log{f_3(z_3,z_4)} \wedge\mathrm{d}\log{f_2(z_2,z_3,z_4)} \wedge \mathrm{d}\log{f_1(z_1,z_2,z_3,z_4)} \,.\label{eq:sunLSE4}
\end{equation}
By generalising the original definition of local integrals provided in~\cite{ArkaniHamed:2010gh},
we can conjecture that an integral in this form is a good candidate for a ``canonical'' integral
defined on an elliptic curve.

We take, therefore, as starting basis $I=\{I_1,I_2,I_3\}$ defined as
\begin{equation}
\label{masters_eqsuninit}
I_1 = I_{0,1,1,0,0}\, , \quad I_2 = I_{1,1,1,0,0} \quad\text{and}\quad I_3 = I_{1,1,2,0,0} \, ,
\end{equation}
where again $I_3$ has been chosen to be proportional to the derivative with respect to the
internal mass $m^2$ of $I_2$.
This basis satisfies the system of differential equations
\begin{align}
\label{masters_eqsun}
     \mathrm d \underline I &= \left(\mathbf{GM}_{m^2}\, \mathrm d m^2 + \mathbf{GM}_{s}\, \mathrm d s \right) \underline I \, , \\
\textbf{GM}_{m^2} &=
\left(
\begin{array}{ccc}
 -\frac{2\eps}{m^2} & 0 & 0 \\
 0 & 0 & 3 \\
 \frac{2\eps^2 s}{m^4 \left(s-m^2\right) \left(s-9 m^2\right)} & \frac{(1+2\eps) (1+3\eps) \left(s-3 m^2\right)}{m^2 \left(s-m^2\right)
   \left(s-9 m^2\right)} & -\frac{s^2-20 m^2 s+27 m^4+\eps \left(s^2-30 m^2 s+45 m^4\right)}{m^2 \left(s-m^2\right) \left(s-9 m^2\right)} \\
\end{array}
\right)\, , \nonumber
\end{align}
where, as before, the Gauss-Manin connection matrix with respect to $s$ follows from a scaling relation.
The tadpole integral $I_1$ is the same as before and remains a suitable candidate for a canonical basis, so we focus our efforts on the top sector. 
Its homogeneous system of differential equations at $\eps=0$ reads
\begin{gleichung}
\label{periodssunrise}
\frac{\partial}{\partial {m^2}}
\left(
\begin{array}{cc}
\mathfrak I_2 \\
\mathfrak I_3 \\
\end{array}
\right) 
= 
\begin{pmatrix}
 0 & 3 \\
 \frac{s-3 m^2}{m^2(s-m^2) (s-9 m^2)} & -\frac{s^2-20m^2 s +27 m^4}{m^2(s-m^2) (s-9 m^2)}
\end{pmatrix}
\left(
\begin{array}{cc}
\mathfrak I_2 \\
\mathfrak I_3 \\
\end{array}
\right)
.
\end{gleichung}
A basis for the solution space for $\mathfrak I_2$ can be constructed explicitly by computing the integral from eq.~\eqref{eq:SunriseMaxCut} over two independent contours dubbed \emph{cycles}. To do so, we consider the kinematic region above the threshold for the production of three massive particles, $s > 9m^2$, such that the branch points of the elliptic curve~\eqref{eq:rootsun} are ordered according to
\begin{equation}
    a_1 < a_2 < a_3 <a_4 \, .
\end{equation}
Two independent cycles can then be chosen as the counter-clockwise contour around the branch cut from $a_1$ to $a_2$ and a similar contour encircling the branch points $a_2$ and $a_3$. The resulting two elliptic functions with a convenient normalisation are given by
\begin{equation}
\label{eq:periodsequalmasssun}
\begin{aligned}
\varpi_0(s,m^2) =&  \frac 2\pi \ \frac{\mathrm{K}\left( k^2 \right)}{\sqrt{(\sqrt{s}-\sqrt{m^2})^3 (\sqrt{s}+3 \sqrt{m^2})}} =\\ 
&= \frac{1}{s} \left[1+3\, \frac{m^2}{s}+15 \left(\frac{m^2}{s}\right)^2+93 \left(\frac{m^2}{s}\right)^3 + \mathcal O\left(\left(\frac{m^2}{s}\right)^4\right)\right] \, , \\
\varpi_1(s,m^2) =&  -\frac 43 \ \frac{\mathrm{K}\left(1-k^2\right)}{\sqrt{(\sqrt{s}-\sqrt{m^2})^3 (\sqrt{s}+3 \sqrt{m^2})}} = \varpi_0(s,m^2) \log \left(\frac{m^2}{s}\right) \\
&\ + \frac{1}{s} \left[4\, \frac{m^2}{s}+26 \left(\frac{m^2}{s}\right)^2+\frac{526}{3} \left(\frac{m^2}{s}\right)^3 + \mathcal O\left(\left(\frac{m^2}{s}\right)^4\right)\right] \, ,
\end{aligned}
\end{equation}
where we have
$$k^2 = \frac {16  \sqrt{s\, m^{2}}^{\, 3}}{(\sqrt{s}-\sqrt{m^2})^3 (\sqrt{s}+3 \sqrt{m^2})}$$
and the complete elliptic integral of the first kind $\mathrm{K}(k^2)$ is defined by
\begin{equation}
\label{def:elliptiK}
    \mathrm{K}(k^2) = \int_{0}^{1} \frac{\mathrm{d} t}{\sqrt{(1-t^2)(1-k^2 t^2)}} \, . 
\end{equation}

Notice that the first solution is holomorphic in a neighbourhood of $m^2/s=0$, while the expansion of the second one contains $\log{(m^2/s)}$. Given these, we can now construct the Wronskian matrix to take the following form
\begin{gleichung}
    \textbf{W} = 
\begin{pmatrix}
 1 & 0 \\ 0 & \frac 13
\end{pmatrix}
\begin{pmatrix}
\varpi_0 & \varpi_1 \\
\partial_{m^2} \varpi_0 & \partial_{m^2} \varpi_1
\end{pmatrix} \, ,
\end{gleichung}
where we suppressed the dependence of the periods on $s,m^2$ for ease of typing.

We proceed by splitting $\mathbf{W}$ into its semi-simple and unipotent part. As outlined before, this splitting is done by demanding the unipotent part to take upper-triangular form with its diagonal normalised to one, while the semi-simple part should be lower-triangular. This yields for the unipotent part
\begin{gleichung}
    \textbf{W}^\text{u} =  \begin{pmatrix}
                                         1 & 2\pi i\, \tau \\ 0 & 1
                                        \end{pmatrix}  \quad\text{with}\quad   \frac{\mathrm d}{\mathrm d\tau} \textbf{W}^\text{u} = \begin{pmatrix}
                                         0 & 2\pi i \\ 0 & 0
                                        \end{pmatrix} \textbf{W}^\text{u} \, ,
\end{gleichung}
where we introduced the $\tau$-parameter defined as
\begin{equation}
\label{tausun}
\begin{aligned}
  \tau = \frac1{2\pi i} \frac{\varpi_1 }{\varpi_0 } =
  \frac1{2\pi i}\left[ \log\left(\frac{m^2}{s }\right) + 4\left(\frac{m^2}{s}\right)+14\left(\frac{m^2}{s}\right)^2+
  \mathcal O \left(\left(\frac{m^2}{s}\right)^3\right) \right] \, .
\end{aligned}
\end{equation}
To write down the semi-simple part, we make use of the \emph{Griffith's transversality conditions}~\cite{MR1288523,MR3965409,MR2451566,Bonisch:2021yfw}, which in the elliptic case correspond to the well-known \emph{Legendre relation} for elliptic integrals,
\begin{equation}
\label{eq:Legendre}
    \varpi_0 \partial_{m^2} \varpi_1 - \varpi_1 \partial_{m^2}  \varpi_0 = \frac{1}{m^2 (s-m^2)(s-9m^2)} \, .
\end{equation}

The left-hand side of~\eqref{eq:Legendre} is proportional to the determinant of $\mathbf{W}$. Consequently, the function on the right-hand side can in practice also be computed conveniently from \emph{Abel's identity} as the solution of a first-order differential equation with rational coefficients. The constant of integration depends on the normalisation of $\varpi_0$ and $\varpi_1$, which is, however, irrelevant to our goal. With this formula at hand, the semi-simple part of $\textbf{W}$ can be written as
\begin{equation}
     \textbf{W}^\text{ss} = 
                \begin{pmatrix} 1 & 0 \\ 0 & \frac 13\end{pmatrix}
                \begin{pmatrix} \varpi_0 & 0 \\ \partial_{m^2} \varpi_0 & \frac{1}{m^2(s-m^2)(s-9m^2)\varpi_0}\end{pmatrix} \, .
\end{equation}

Once we reach this point, the remaining steps work as in the polylogarithmic case considered in the previous subsection. After rotating the top sector with the inverse of the semi-simple part,
we perform a simple rescaling and integrate out a total derivative to arrive at $\eps$-factorised differential equations for the equal-mass sunrise graph:
\begin{gleichung}
    \frac{\partial}{\partial m^2}\, \underline J &= \eps\, \textbf{GM}_{m^2}^{\eps}\, \underline J \quad\text{with}\quad \underline J = \mathbf{T}\, \underline I \, , \\
\label{eq_rotsunrise}
       \mathbf{T} &=  
          \begin{pmatrix} 1 & 0 & 0\\ 0 & 1 & 0 \\ 0 & \frac{s^2-30 m^2 s+45 m^4}{2}\varpi_0^2 & 1\end{pmatrix}
          \begin{pmatrix} \eps^2 & 0 & 0\\ 0 & \eps^2 & 0 \\ 0 & 0 & \eps \end{pmatrix}
          \begin{pmatrix}
          1 & \begin{matrix} 0 \ & \ 0 \end{matrix} \\
          \begin{matrix} 0 \\ 0 \end{matrix} & (\textbf{W}^\text{ss})^{-1}
          \end{pmatrix} \, , \\
\textbf{GM}_{m^2}^{\eps} &= 
\left(
\begin{array}{ccc}
 -\frac{2}{m^2} & 0 & 0 \\
 0 & -\frac{s^2-30 m^2 s+45 m^4}{2 m^2 (s-m^2) (s-9 m^2)} & \frac{1}{m^2 (s-m^2) (s-9 m^2) } \varpi_0^2\\
 \frac{6 s }{m^2} \varpi_0 & \frac{(3 m^2+s)^4 }{4 m^2 (s-m^2) (s-9 m^2)} \frac{1}{\varpi_0^2} & -\frac{s^2-30 m^2 s+45 m^4}{2
   m^2 (s-m^2) (s-9 m^2)} \\
\end{array}
\right) \, .
\end{gleichung}
It can be verified that the basis $\underline J$ is, up to a simple constant rotation, 
the same basis as derived with other methods in~\cite{Adams:2018yfj,Dlapa:2022wdu}. 
The rotation $\mathbf{T}$ in \eqref{eq_rotsunrise} is constructed from functions with rational dependence on $s$, $m^2$, $\varpi_0$ and $\partial_{m^2} \varpi_0$. Notice, however, that the final $\eps$-factorised Gauss-Manin connection matrix in \eqref{eq_rotsunrise} only contains $\varpi_0$ and \emph{not} its derivative. Moreover, all of its entries have at most simple poles in all singular limits, because $\varpi_0 \sim 1/m^2$ as $m^2 \rightarrow \infty$. If desired, it is also possible to change variables to the canonical variable $\tau$ defined in \eqref{tausun}. Then the entries of $\textbf{GM}_{m^2}^{\eps}$ will be given by modular forms of the corresponding monodromy group of the sunrise $\Gamma_1(6)$, as it was first observed in~\cite{Bloch:2013tra,adams2018feynman}.
On a side note, we would like to mention that upon replacing $\varpi_0$ by \emph{any} $\mathbb{Q}$-linear combination of $\varpi_0$ and $\varpi_1$ in $\mathbf{T}$ would lead to the identical $\eps$-factorised Gauss-Manin connection matrix $\eps \, \textbf{GM}_{m^2}^{\eps}$ with $\varpi_0$ replaced by the same linear combination.

Finally, we could have started from a different basis, where the second master integral
in the top sector could be chosen by embedding the sunrise in the kite integral family,
and selecting as independent a reducible integral with an extra massless propagator.
Starting from this different basis, one can prove that our procedure produces an
$\epsilon$-factorised basis that is identical to the one obtained here, modulo a rotation by a constant numerical matrix.

\subsection{The sunrise with three non-vanishing masses of which two are equal}
As a third introductory example, we consider now the fully massive case, where two of the three internal masses are equal but different from the third mass. This increases the number of scales in the problem by one compared to before and gives rise to a second independent tadpole master integral, as well as to a third master integral in the top sector. 
As initial basis we take the master integrals $\underline I=\{ I_1, \hdots, I_5\}$ given by
\begin{equation}
\label{masters_twosun}
I_1 = I_{1,1,0,0,0}\,, \quad I_2 = I_{0,1,1,0,0}\,, \quad I_3 = I_{1,1,1,0,0}\,, \quad I_4=I_{1,1,2,0,0}\,, \quad I_5=I_{1,1,1,-1,0} \, .
\end{equation}
They satisfy a system of differential equations of the form
\begin{equation}
\label{gmsystem_sunrise}
     \mathrm d \underline I = \left( \mathbf{GM}_{s}\, \mathrm d s +\mathbf{GM}_{m_1^2}\, \mathrm d m_1^2 + \mathbf{GM}_{m_2^2}\, \mathrm d m_2^2 \right) \underline I \, .
\end{equation}
The explicit expressions for the Gauss-Manin connections exceed what can be printed in a readable format, but they can be found in an ancillary file. 

Let us now comment on how we chose the starting basis in eq.~\eqref{masters_twosun}.  
The tadpoles are trivial, while we know from the previous example how to choose the first and second integral in the sunrise top sector. From the maximal cut in eq.~\eqref{eq:SunriseMaxCut}, we know that
the geometry remains elliptic in even numbers of dimensions. Equivalently, the Picard-Fuchs operator associated to the sunrise sector factorises at $\eps=0$ into a second-order and a first-order irreducible operator. We expect, therefore, that there should be
a minimally coupled system of two master integrals  in $d=2$, and not one of three master integrals.

To make this decoupling manifest, the integral $I_5$ was chosen by analysing the integrand associated to the maximal cut in a loop-by-loop Baikov representation as in~\eqref{eq:sunriseBaikov} and realising that when the masses are different, one can choose an independent candidate which possesses a non-vanishing, constant residue at $z_4=\infty$. Explicitly, we have
\begin{equation}
\label{eq:SunriseMaxCutNumInt}
    \text{MaxCut}(I_{1,1,1,-1,0}) \sim \int \frac{\mathrm{d}z_4 \, z_4}{\sqrt{P_4(z_4)}} \sim \int \frac{\mathrm{d}y}{y \sqrt{\widetilde{P}_4(y)}}   \quad \mbox{with} \quad  \widetilde{P}_4(y) = y^4 \, P_4(1/y) \,.
\end{equation}
Similarly, one could decouple the third integral using a Gram determinant as it was shown in~\cite{Remiddi:2013joa}.
The maximal cut system of the $(2 \times 2)$-block formed by the master integrals $I_3$ and $I_4$ encodes the elliptic part of the sector for which we want to compute the semi-simple part of the Wronskian matrix. This works exactly as in the equal-mass case. Nevertheless, we will see something new happening in this case. Let us quickly go through the steps involved. 
Explicitly, the maximal cut system reads
\begin{gleichung}
\mathrm d \begin{pmatrix} \mathfrak I_3 \\ \mathfrak I_4 \end{pmatrix} &= \left(  \mathbf{GM}_{s}^{\mathcal E}\, \mathrm d s  + \mathbf{GM}_{m_1^2}^{\mathcal E}\, \mathrm d m_1^2 + \mathbf{GM}_{m_2^2}^{\mathcal E}\, \mathrm d m_2^2 \right)\begin{pmatrix} \mathfrak I_3 \\ \mathfrak I_4 \end{pmatrix} \quad\text{with} \\
\mathbf{GM}_{s}^{\mathcal E} &= 
\left(
\begin{array}{cc}
 -\frac{s-m_2^2}{s \left(s-m_1^2\right)} & -\frac{m_2^2 \left(3 s+m_1^2-4 m_2^2\right)}{s \left(s-m_1^2\right)} \\
 -\frac{s \left(s+3 m_1^2\right)-\left(7 s+m_1^2\right) m_2^2+4 m_2^4}{s \left(s-m_1^2\right) \Delta } & \frac{s-m_2^2}{s \left(s-m_1^2\right)}+\frac{2
   \left(-s+m_1^2+4 m_2^2\right)}{\Delta } \\
\end{array}
\right) \, , \\
\mathbf{GM}_{m_1^2}^{\mathcal E} &= 
\left(
\begin{array}{cc}
 \frac{m_1^2-m_2^2}{m_1^2 \left(s-m_1^2\right)} & \frac{m_2^2 \left(s+3 m_1^2-4 m_2^2\right)}{m_1^2 \left(s-m_1^2\right)} \\
 \frac{m_1^2 \left(3 s+m_1^2\right)-\left(s+7 m_1^2\right) m_2^2+4 m_2^4}{m_1^2 \left(s-m_1^2\right) \Delta} & -\frac{m_1^2-m_2^2}{m_1^2
   \left(s-m_1^2\right)}+\frac{2 \left(s-m_1^2+4 m_2^2\right)}{\Delta} \\
\end{array}
\right) \, ,\\
\mathbf{GM}_{m_2^2}^{\mathcal E} &= 
\left(
\begin{array}{cc}
 0 & 2 \\
 \frac{s+m_1^2-6 m_2^2}{m_2^2 \, \Delta} & \frac{8 \left(s+m_1^2-4 m_2^2\right)}{\Delta} -\frac{1}{m_2^2}\\
\end{array}
\right) \, , 
\end{gleichung}
where
\begin{equation}
    \Delta \coloneqq \Delta (s,m_1^2,m_2^2) = \left(s-m_1^2\right)^2 - 8 m_2^2 \left(s+m_1^2- 2m_2^2\right) \, .
\end{equation}
In the kinematic region above the threshold for the production of three massive particles, $s>(m_1+2m_2)^2$, the elliptic functions spanning the solution space can be written as
\begin{equation}
\label{eq:periodssunmultivariatecase}
\begin{aligned}
\varpi_0(s,m_1^2,m_2^2) &= \frac 2\pi \ \dfrac{\mathrm{K}\left( k^2 \right)}{\sqrt{Z(s,m_1^2,m_2^2)}}   \\
&= \frac{1}{s} \left[1+\frac{m_1^2+2m_2^2}{s}+\frac{m_1^4 + 8m_1^2 m_2^2 + 6m_2^4}{s^2}+ \mathcal O\left(\left(\frac{m_i^2}{s}\right)^3\right)\right] \, , \\
\varpi_1(s,m_1^2,m_2^2) &=  -4 \ \dfrac{\mathrm{K}\left(1- k^2 \right)}{\sqrt{Z(s,m_1^2,m_2^2)}} = \varpi_0 \left[\log\left( \frac{m_1^2}{s} \right) + 2 \log\left( \frac{m_2^2}{s} \right)\right]\\
    & \quad+ \frac{1}{s} \left[\frac{4m_1^2+8m_2^2}{s}+\frac{6m_1^4 + 40 m_1^2 m_2^2 + 32 m_2^4}{s^2}+ \mathcal O\left(\left(\frac{m_i^2}{s}\right)^3\right)\right] \, , 
\end{aligned}
\end{equation}
where
\begin{gleichung}
\label{eq:k2sun2}
    k^2 &= 16\dfrac { \sqrt{s\, m_1^2}\, m_2^2}{Z(s,m_1^2,m_2^2)} \quad\text{and}\quad
    Z(s,m_1^2,m_2^2) = \left(s-m_1^2\right){}^2-4 m_2^2 \left(\sqrt{s} - \sqrt{m_1^2}\right)^2 \, .
\end{gleichung}
We will again omit the explicit dependence of the periods on the kinematic variables whenever convenient. The resulting Wronskian matrix takes the form 
\begin{gleichung}
    \textbf{W} &= 
    \begin{pmatrix}
        1 & 0 \\
        0 & \frac{1}{2}
    \end{pmatrix} 
    \begin{pmatrix}
        \varpi_0 & \varpi_1 \\
        \partial_{m_2^2}\varpi_0 & \partial_{m_2^2}\varpi_1
    \end{pmatrix} \quad\text{with} \\
\mathrm d \textbf{W} &= \left( \mathbf{GM}_{s}^{\mathcal E}\, \mathrm d s + \mathbf{GM}_{m_1^2}^{\mathcal E}\, \mathrm d m_1^2 + \mathbf{GM}_{m_2^2}^{\mathcal E}\, \mathrm d m_2^2  \right)
\textbf{W} \, .
\end{gleichung}
The unipotent piece is analogous to the equal-mass case but we will not need its explicit form in the following. What we do need is the semi-simple part which, after employing the Legendre relation, reads
\begin{equation}
    \textbf{W}^\text{ss} = 
    \begin{pmatrix}
        1 & 0 \\
        0 & \frac{1}{2}
    \end{pmatrix} 
    \begin{pmatrix}
        \varpi_0 & 0 \\
        \partial_{m_2^2}\varpi_0 & \frac{2}{m_2^2 \, \Delta \, \varpi_0}
    \end{pmatrix} \, .
\end{equation}

With this ingredient, we can now construct the following rotation to apply to our initial basis $\underline I$ to reach upper-triangular $\eps$-form:
\begin{gleichung} 
\label{rotsun1}
   \tilde{\mathbf{T}} =  
          \begin{pmatrix}  \eps^2 & 0 & 0 & 0 & 0 \\
                           0 &   \eps^2 & 0 & 0 & 0 \\
                           0 & 0 &   \eps^2 & 0 & 0 \\
                           0 & 0 & 0 &   0 & \eps^2 \\
                           0 & 0 & 0 & \eps & 0\end{pmatrix} 
         \begin{pmatrix}  1 & 0 & 0 & 0 & 0 \\
                           0 & 1 & 0 & 0 & 0 \\
                           0 & 0 &\varpi_0 & 0 & 0 \\
                           0 & 0 &\frac12\partial_{m_2^2}\varpi_0  & \frac{1}{m_2^2 \, \Delta \, \varpi_0}& 0 \\
                           0 & 0 & 0 & 0 & 1\end{pmatrix}^{-1} \, .
\end{gleichung}
Notice that this step includes swapping the fourth and fifth integral in the basis after 
the rotation coming from the semi-simple part of the Wronskian matrix. 
Finally, we want to integrate out the remaining non-$\eps$-factorised terms. 
Here, however, one immediately realises that, in contrast to the equal-mass case, they are not just total derivatives of rational functions in $s, m_1^2, m_2^2$, $\varpi_0$ and $\partial_{m_2^2}\varpi_0$. 
To integrate them out, we need to introduce a new function, which we call $G \coloneqq G(s,m_1^2,m_2^2)$. The new function $G$ can be defined by its partial derivatives
\begin{equation}
\label{def:Gsunrise}
\begin{aligned}
      \partial_{s} G &= -\frac{m_2^2 \left(3 s+m_1^2-4 m_2^2\right)}{s \left(s-m_1^2\right)}\, \varpi_0 +\frac{m_2^2 \, \Delta }{2 s \left(s-m_1^2\right)} \, \partial_{m_2^2}\varpi_0 \, , \\
      \partial_{m_1^2} G &= \frac{m_2^2 \left(s+3 m_1^2-4 m_2^2\right)}{m_1^2 \left(s-m_1^2\right)}\, \varpi_0 -\frac{m_2^2 \, \Delta }{2 m_1^2 \left(s-m_1^2\right)} \, \partial_{m_2^2}\varpi_0 \, , \\
      \partial_{m_2^2} G &= 2 \, \varpi_0 \, .
\end{aligned}
\end{equation}
Using the differential equations for $\varpi_0$, it is easy to verify the Schwarz integrability conditions
\begin{equation}
\label{eq:SchwarzintcondSunrise}
\begin{aligned}
    \partial_{s} \partial_{m_1^2} G &= \partial_{m_1^2} \partial_{s} G \, , \quad
    \partial_{s} \partial_{m_2^2} G &= \partial_{m_2^2} \partial_{s} G \, , \quad
    \partial_{m_1^2} \partial_{m_2^2} G &= \partial_{m_2^2} \partial_{m_1^2} G \, .
\end{aligned}
\end{equation}
The last line in eq.~\eqref{def:Gsunrise} can be solved to express $G$ as an integral over $\varpi_0$,
\begin{equation}
\label{eq_Gsunrise}
     G(s,m_1^2,m_2^2)  = 2 \int_0^{m_2^2} \mathrm{d}u \ \varpi_0(s,m_1^2,u)\, .
\end{equation} 
The lower boundary was set to zero for convenience. We stress, however, that the particular choice of integration constant won't impact the factorisation of the differential equations. Similarly to~\eqref{eq:SchwarzintcondSunrise}, by using the differential equations for $\varpi_0$ under the integral sign,  one can show explicitly  that \eqref{eq_Gsunrise} also satisfies the other two relations in~\eqref{def:Gsunrise}. 

By using the explicit representation for $\varpi_0$ from eq.~\eqref{eq:periodssunmultivariatecase} in terms of an elliptic integral of the first kind, (c.f. eq.~\eqref{def:elliptiK}) we can exchange the order
of integration and derive a representation for $G$ in terms of an elliptic integral of the third kind
\begin{equation}
\label{eq:Gasellipticint}
\begin{aligned}
    G(s,m_1^2,m_2^2) &= \frac{4}{\pi} \int_{0}^{1}\mathrm{d} t \int_0^{m_2^2} \frac{\mathrm{d}u}{\sqrt{Z(s,m_1^2,u)}\sqrt{\left(1-t^2\right)\left(1-k^2  \, t^2\right)}}  \\
    &= -\frac{2 \left(\sqrt{s}+\sqrt{m_1^2}\right)^2}{\pi  \sqrt{Z(s,m_1^2,m_2^2)}} \, \mathrm{\Pi} \left(a, k^2\right) +4 m_2^2 \, \varpi_0 +1   \,.
\end{aligned}
\end{equation}
Notice that $k^2$ was defined in eq.~\eqref{eq:k2sun2} 
and in the integral sign, we renamed $m_2^2=u$ and
\begin{equation}
a=-\frac{4 \sqrt{s \, m_1^2}}{\left(\sqrt{s}-\sqrt{m_1^2}\right)^2}\,.
\end{equation}
We recall here that the elliptic integral of the third kind $\mathrm{\Pi}(n,k^2)$ is defined as 
\begin{equation}
\label{def:elliptiPi}
    \mathrm{\Pi}(n,k^2) = \int_{0}^{1} \frac{\mathrm{d} t}{\left(1- n \, t^2 \right)\sqrt{(1-t^2)(1-k^2 t^2)}} \, .
\end{equation}
That $G$ admits a representation in terms of an elliptic integral of the third kind is not at all surprising. In fact, if we solved the homogeneous system at $\eps=0$ for the whole $3 \times 3$ sunrise sector, the corresponding Wronskian matrix would contain terms related to integrating the maximal cut of $I_5$ (c.f. eq. \eqref{eq:SunriseMaxCutNumInt}) over the cycles of the elliptic curve. Its integrand corresponds to the differential of the third kind on an elliptic curve and its integration yields elliptic integrals of the third kind, as can be verified readily by an explicit computation. We expect therefore functions of this type to appear in the Laurent expansion of the sunrise master integrals to arbitrary orders in $\eps$. In particular, one can verify that
\begin{equation}
    \text{MaxCut}(I_{1,1,1,-1,0}) \propto \left(4 m_2^2 \varpi_0 -G \right) \, ,
\end{equation}
when integrated over the cycle of the elliptic curve for which $\text{MaxCut}(I_{1,1,1,0,0}) \propto  \varpi_0$. 

After introducing $G$, the factorisation of $\eps$ is achieved as follows:
\begin{gleichung}
\label{eq_newGsunrise}
    \mathrm d \underline J &= \eps \left( \textbf{GM}_{s}^{\eps} \mathrm d s  + \textbf{GM}_{m_1^2}^{\eps} \mathrm d m_1^2 + \textbf{GM}_{m_2^2}^{\eps} \mathrm d m_2^2 \right) \underline J \quad\text{with}\quad \underline J = \mathbf{T}\, \underline I \, , \\
     \mathbf{T} &=  \begin{pmatrix} 1 & 0 & 0 & 0 & 0 \\
 0 & 1 & 0 & 0 & 0 \\
 0 & 0 & 1 & 0 & 0 \\
 0 & 0 & G & 1 & 0 \\
 -G & -\frac{1}{2} \, G & \frac{3}{4} \, G^2 & \frac{3}{2}
  \, G & 1 \end{pmatrix}
   \begin{pmatrix}  1 & 0 & 0 & 0 & 0 \\
                           0 & 1 & 0 & 0 & 0 \\
                           0 & 0 & 1 & 0 & 0 \\
                           0 & 0 & -4 m_2^2 \, \varpi_0 & 1 & 0 \\
                           0 & 0 &\alpha \, \varpi_0^2 & 0 & 1
        \end{pmatrix}
   \tilde{\mathbf{T}} \, , \\
    \alpha &= \frac14\left(\Delta -16m_2^2 \left(s+m_1^2-4 m_2^2\right)\right)\,.
\end{gleichung}
The terms in eq.~\eqref{eq_newGsunrise} that are independent of $G$ are required to integrate out all terms containing $\partial_{m_2^2}\varpi_0$ in the Gauss-Manin connection matrix for the variable $m_2^2$. After this step, there are remaining terms below its diagonal, that are not yet proportional to $\eps$, but that depend on $\varpi_0$. Hence, we need to introduce a function whose derivative is proportional to $\varpi_0$, from which $G$ emerges naturally. The $\eps$-independent Gauss-Manin connection matrices $\textbf{GM}_{z}^{\eps}$ for $z=s, m_1^2, m_2^2$ read explicitly as follows:
\begin{gleichung}
    \textbf{GM}^{\eps}_{z} &= 
    \begin{pmatrix}
        \begin{matrix}
            -A_4^z-A_5^z & 0 \\
            0 & -2 A_5^z \\
            2 A_2^z & A_2^z \\
            A_3^z+A_4^z-2 A_5^z+2 A_7^z & -\frac{1}{2}A_3^z+\frac{1}{2}A_4^z+A_7^z \\
            \frac{3}{2} A_8^z-2 A_9^z-2 A_{10}^z & \frac{3}{4} A_8^z+2 A_{10}^z 
        \end{matrix} & 
        \begin{matrix}
            0 & 0 & 0 \\
            0 & 0 & 0 \\
            \\
            \left. \right. & \textbf{GM}^{\text{s},\eps}_{z} \\
        \end{matrix}
    \end{pmatrix}\, , \\
    \textbf{GM}^{\text{s},\eps}_{z} &=
    \begin{pmatrix}
         \tilde{A} & -3 A_2^z & 2 A_1^z \\
        \frac{3}{2} A_8^z & -\frac{1}{2}A_3^z-\frac{3}{2} A_4^z-3 A_7^z & 2 A_2^z \\
        \frac{9}{8}A_{11}^z & -\frac{9}{4} A_8^z & \tilde{A}
    \end{pmatrix} \, , \\
    \tilde{A} &= \frac{5}{4} A_3^z+\frac{1}{4}A_4^z-\frac{1}{2}A_5^z-A_6^z+\frac{3}{2} A_7^z \, ,
\end{gleichung}
where we have
\allowdisplaybreaks
\begin{align}
    &A_1^{s} = -\frac{3 s+m_1^2-4 m_2^2}{2 s \left(s-m_1^2\right) \Delta \, \varpi_0^2} \, , \quad A_1^{m_1^2} = A_1^{s}\vert_{s\leftrightarrow m_1^2} \, , \quad A_1^{m_2^2} = \frac{1}{m_2^2 \, \Delta \, \varpi_0^2} \, , \nonumber \\
    &A_2^{s} = -\frac{1}{2s \left(s-m_1^2\right) \varpi_0}-\frac{\left(3 s+m_1^2-4 m_2^2\right) G}{2s \left(s-m_1^2\right) \Delta \, \varpi_0^2} \, , \quad A_2^{m_1^2} = A_2^{s}\vert_{s\leftrightarrow m_1^2} \, , \quad A_2^{m_2^2} = \frac{G}{m_2^2 \, \Delta \, \varpi_0^2} \, , \nonumber \\ 
    &A_3^{s} = \frac{1}{s} \, , \quad A_4^{m_1^2} = \frac{1}{m_1^2} \, ,\quad A_5^{m_2^2} = \frac{1}{m_2^2}  \, , \quad A_3^{m_1^2} = A_3^{m_2^2}=A_4^{s} =A_4^{m_2^2} =A_5^{s} = A_5^{m_1^2} = 0 \, ,  \nonumber \\
    &A_6^{s} = \frac{1}{s-m_1^2}+\frac{\partial_{s} \Delta}{\Delta} \, , \quad A_6^{m_1^2} = A_6^{s}\vert_{s\leftrightarrow m_1^2} \, , \quad A_6^{m_2^2} = \frac{\partial_{m_2^2} \Delta}{\Delta} \, , \nonumber \\
    &A_7^{s} = \, \frac{2m_2^2}{s \left(s-m_1^2\right)}-\frac{G}{s \left(s-m_1^2\right) \varpi_0 }-\frac{\left(3 s+m_1^2-4 m_2^2\right) G^2}{2 s \left(s-m_1^2\right) \Delta \, \varpi_0 ^2} \, ,  \nonumber \\
    &A_7^{m_1^2} = A_7^{s}\vert_{s\leftrightarrow m_1^2} \, , \quad A_7^{m_2^2} = \, \frac{G^2}{m_2^2 \Delta \, \varpi_0 ^2} \, , \nonumber \\
    &A_8^{s} =  -\frac{\left[\left(s-m_1^2\right)^2-16m_2^2\left(m_1^2-3m_2^2\right)\right] \varpi_0 }{6 s \left(s-m_1^2\right)} +\frac{P_1(s,m_1^2,m_2^2) \, G}{6 s \left(s-m_1^2\right) \Delta} \nonumber \\ 
    &\quad \quad - \frac{3 G^2}{2 s \left(s-m_1^2\right) \varpi_0 }  -\frac{\left(3 s+m_1^2-4 m_2^2\right) G^3}{2 s \left(s-m_1^2\right) \Delta \, \varpi_0^2} \, , \nonumber \\
    &A_8^{m_1^2} = A_8^{s}\vert_{s\leftrightarrow m_1^2} \, , \quad A_8^{m_2^2} = \, \frac{8 \varpi_0}{3} -\frac{\alpha \, G}{3 m_2^2 \, \Delta }+\frac{G^3}{m_2^2 \, \Delta \, \varpi_0 ^2} \, ,  \nonumber \\ 
    &A_{9}^{s} = \varpi_0 \, , \quad A_9^{m_1^2} = -\frac{\left(s+m_1^2-4 m_2^2\right) \varpi_0 }{2 m_1^2}-\frac{G}{2 m_1^2} \, , \quad A_{9}^{m_2^2} = \frac{\left(-s+m_1^2-4 m_2^2\right) \varpi_0 }{2 m_2^2}+\frac{G}{2 m_2^2} \, ,  \nonumber \\
    &A_{10}^{s} = A_9^{m_1^2}\vert_{s\leftrightarrow m_1^2} \, , \quad A_{10}^{m_1^2} = A_9^{s}\vert_{s\leftrightarrow m_1^2} \, , \quad A_{10}^{m_2^2} = A_9^{m_2^2}\vert_{s\leftrightarrow m_1^2} \, , \nonumber \\ 
    & \nonumber \\
    &A_{11}^{s} = \, \frac{P_2(s,m_1^2,m_2^2) \, \varpi_0 ^2}{18 s \left(s-m_1^2\right) \Delta }-\frac{2 \left[\left(s-m_1^2\right){}^2-16 m_2^2 \left(m_1^2-3 m_2^2\right)\right] \varpi_0 \,  G}{3 s
   \left(s-m_1^2\right)} \nonumber \\ 
    &\quad \quad +\frac{P_1(s,m_1^2,m_2^2) \, G^2}{3 s \left(s-m_1^2\right) \Delta }-\frac{2 G^3}{s \left(s-m_1^2\right) \varpi_0 }-\frac{\left(3 s+m_1^2-4
   m_2^2\right) G^4}{2 s \left(s-m_1^2\right) \Delta \, \varpi_0 ^2} \, , \nonumber \\
    &A_{11}^{m_1^2} = A_{11}^{s}\vert_{s\leftrightarrow m_1^2} \, , \quad A_{11}^{m_2^2} = \frac{P_3(s,m_1^2,m_2^2) \, \varpi_0 ^2}{9 m_2^2 \, \Delta }+\frac{32 \varpi_0 \, G}{3}-\frac{2 \alpha \, G^2}{3 m_2^2 \, \Delta }+\frac{G^4}{m_2^2 \, \Delta \, \omega ^2} \, .\nonumber 
\end{align}
The $P_i(s,m_1^2,m_2^2)$ for $i=1,2,3$, are polynomials whose explicit expressions read
\begin{align}
    P_1(s,m_1^2,m_2^2) =& -\left(s-m_1^2\right){}^2 \left(5 s+7 m_1^2\right)+\left(44 s^2-72 s m_1^2+92 m_1^4\right) m_2^2  \\
    &-80 \left(3 s+5 m_1^2\right) m_2^4+576 m_2^6 \nonumber \\
    P_2(s,m_1^2,m_2^2) =& -\left(s-m_1^2\right){}^4 \left(3 s+m_1^2\right)+4 \left(s-m_1^2\right){}^2 \left(s^2-42 s m_1^2+9 m_1^4\right) m_2^2 \nonumber \\
    &-32 \left(s^3+37 s^2 m_1^2-21 s m_1^4+15 m_1^6\right) m_2^4 \nonumber \\
    &+128 \left(11 s^2+22 s m_1^2+23 m_1^4\right) m_2^6-768 \left(9 s+11 m_1^2\right) m_2^8+9216 m_2^{10} \nonumber \\
    P_3(s,m_1^2,m_2^2) =& \left(s-m_1^2\right){}^4+16 \left(s-m_1^2\right){}^2 \left(s+m_1^2\right) m_2^2 \nonumber \\
    &-32 \left(7 s^2-22 s m_1^2+7 m_1^4\right) m_2^4+768 \left(s+m_1^2\right) m_2^6-768 m_2^8 \, . \nonumber
\end{align}
We would like to point out that, if we rescale the last integral in the basis by a factor $-2/3$, the entries of $\textbf{GM}^{\text{s},\eps}_{z} $ become persymmetric. Notice that, as in the previous case, none of the kernels in the Gauss-Manin~\eqref{eq_newGsunrise} contains derivatives of the period $\varpi_0$. While this allows us to exclude obvious integration by parts identities 
among the integration kernels, non-trivial relations cannot be ruled out and should be investigated explicitly case by case. 

It is interesting to study how the basis $\underline J$ yielded by our procedure behaves in the limits $m_1 \rightarrow 0$ and $m_1 \rightarrow m_2$, i.e. how it is related to the $\eps$-factorised bases from the previous two subsections. In both cases, the integrals $I_1$ and $I_5$ in our starting basis can be written as linear combinations of $I_2, I_3$ and $I_4$. Moreover, for $s>9\tilde{m}^2>0$ one can prove the following functional relations 
\begin{align}
    &\varpi_0(s,0,\tilde{m}^2) = \frac{1}{r(s,\tilde{m}^2)} \, , \ &&\varpi_0(s,\tilde{m}^2,\tilde{m}^2) = \varpi_0(s,\tilde{m}^2) \, , \nonumber \\ 
    &\partial_{m_2^2} \varpi_0(s,0,\tilde{m}^2) = \frac{2s}{ r(s,\tilde{m}^2)^3} \, , \ &&\partial_{m_2^2} \varpi_0(s,\tilde{m}^2,\tilde{m}^2) = \frac{2}{3} \partial_{m^2} \varpi_0(s,\tilde{m}^2) \, , \\ 
    &G(s,0,\tilde{m}^2) = 1- \frac{r(s,\tilde{m}^2)}{s} \, ,\ &&G(s,\tilde{m}^2,\tilde{m}^2) = \frac{9\tilde{m}^2 -s}{3}\varpi_0(s,\tilde{m}^2)+\frac{1}{3} \, . \nonumber
\end{align}
The objects $r(\cdot,\cdot)$ and $\varpi_0(\cdot,\cdot)$ on the right-hand-sides have been defined in eqs.~\eqref{eq:sqrtsun} and~\eqref{eq:periodsequalmasssun}. The easiest way to verify these relations is from the explicit representations for $\varpi_0$ and $G$ in terms of elliptic integrals in eq.~\eqref{eq:periodssunmultivariatecase} and eq.~\eqref{eq:Gasellipticint}, respectively. Using the above relations, it can be checked that all the integrals in the basis $\underline J$ reduce to simple $\mathbb{Q}$-linear combinations of the $\eps$-factorised bases we found in the previous two subsections, c.f. eq.~\eqref{def:cbasissunPLus} and eq.~\eqref{eq_rotsunrise}.

Before we conclude this section let us comment on the case of three different internal masses. This setup gives rise to a third independent tadpole, but more importantly, a fourth master in the top sector, for which a convenient choice of initial basis integral is given by $I_{1,1,1,0,-1}$. The rotation to an $\eps$-factorised basis  is constructed in an analogous fashion as in eqs.\eqref{rotsun1} and \eqref{eq_newGsunrise}. However, as expected, one now has to introduce two new functions in the last step to integrate out all undesired terms, and which correspond to two differentials of the third kind. These can again be represented as integrals over the corresponding holomorphic period but with additional rational functions in the integrand. The explicit expressions are provided in an ancillary file to this manuscript.

%% file: Sec3_Applications.tex
% !TEX encoding = UTF-8 Unicode

\section{Applications to Feynman integrals with a single elliptic curve}
\label{sec:applications}

In the previous section, we have worked out explicitly various mass configurations for the two-loop sunrise graph, showing how our procedure allows us to obtain $\eps$-factorised
systems of differential equations in almost algorithmic steps. For the multi-scale cases, this came at the cost of introducing extra objects, defined as integrals over the holomorphic period of the elliptic curve and rational functions. We have seen that these can be related to differential forms of the third kind.
In this section, we will present several examples of single- and multivariate Feynman integral families related to a single elliptic curve, where our approach can be successfully applied. We will see that the patterns we observed can be extended also to three- and four-point functions.

\subsection{A single-scale elliptic three-point function}
\label{ssec:triagnle1}

%-----Picture Triangle1 Graph-------------
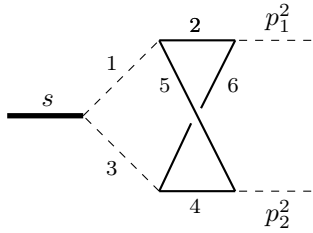
\begin{figure}[!h]
\centering
\begin{tikzpicture}
\coordinate (links) at (-0.5,0);
\coordinate (mitte) at (0.5,0);
\coordinate (mmitte) at (1.92,-0.1); 
\coordinate (mmmitte) at (2.05,0.1);
\coordinate (oben) at (1.5,1);
\coordinate (unten) at (1.5,-1);
\coordinate (obenr) at (2.5,1);
\coordinate (untenr) at (2.5,-1);
\coordinate (obenrr) at (3.5,1);
\coordinate (untenrr) at (3.5,-1);
\begin{scope}
\draw [-, thick,postaction={decorate}] (oben) to [bend right=0]  (obenr);
\draw [-, thick,postaction={decorate}] (untenr) to [bend right=0]  (unten);
\draw [-, thick,postaction={decorate}] (oben) to [bend right=0]  (untenr);
\draw [-, thick,postaction={decorate}] (obenr) to [bend right=0]  (mmmitte);
\draw [-, thick,postaction={decorate}] (unten) to [bend right=0]  (mmitte);
\end{scope}
\begin{scope}
\draw [-, dashed,postaction={decorate}] (obenrr) to [bend right=0]  (obenr);
\draw [-, dashed,postaction={decorate}] (untenrr) to [bend right=0]  (untenr);
\draw [-, dashed,postaction={decorate}] (mitte) to [bend right=0]  (oben);
\draw [-, dashed,postaction={decorate}] (unten) to [bend right=0]  (mitte);
\end{scope}
\begin{scope}
\draw [-, line width=2pt,postaction={decorate}] (mitte) to [bend right=0]  (links);
\end{scope}
\node (d1) at (0.9,0.7) [font=\scriptsize, text width=.2 cm]{1};
\node (d2) at (0.9,-0.7) [font=\scriptsize, text width=.2 cm]{3};
\node (d3) at (2,1.2) [font=\scriptsize, text width=.2 cm]{2};
\node (d4) at (2,-1.2) [font=\scriptsize, text width=.2 cm]{4};
\node (d5) at (2,1.2) [font=\scriptsize, text width=.2 cm]{2};
\node (d6) at (1.6,0.4) [font=\scriptsize, text width=.2 cm]{5};
\node (d7) at (2.5,0.4) [font=\scriptsize, text width=.2 cm]{6};
\node (d8) at (0.05,0.2) [font=\small, text width=.2 cm]{$s$};
\node (d9) at (3,1.3) [font=\small, text width=.2 cm]{$p_1^2$};
\node (d10) at (3,-1.3) [font=\small, text width=.2 cm]{$p_2^2$};
\end{tikzpicture}
\caption{The graph of non-planar triangle number 1.}
\label{fig:triangle1}
\end{figure}
%------------------------

\noindent
As the first example beyond the sunrise topology, we consider the triangle graph shown in figure \ref{fig:triangle1}, for which we use the following integral family~\cite{vonManteuffel:2017hms,Broedel:2019hyg}\footnote{We thank Xing Wang for having shared his results for this integral family with us prior to publication~\cite{Xing:2023}.}
\begin{gleichung}
\label{def_triangle1family}
D_1 &= (k_1-p_1)^2\,, \quad &D_2 &= (k_2-p_1)^2-m^2\,, \quad  &D_3 &= (k_1+p_2)^2\,, \\
D_4 &= (k_1-k_2+p_2)^2-m^2\,, \quad &D_5 &= (k_1-k_2)^2-m^2\,, \quad  &D_6 &= k_2^2-m^2\,, \\
N_1 &= k_1^2 \, .
\end{gleichung}
We take $p_1^2=p_2^2=0$ and $s = (p_1+p_2)^2$ such that the integrals, suitably normalised, depend on the single variable 
\begin{equation}
    z=-\frac{m^2}{s} \, .
\end{equation}
At variance to the sunrise, we consider this two-loop triangle in $d=4-2\eps$ dimensions and we also set $m =1$ for simplicity.

Upon IBP reduction, we find that the problem has $11$ master integrals. Following our general prescription, a suitable starting basis $\underline I=\{ I_1,\hdots, I_{11} \}$ is given by the integrals
\begin{gleichung}
\label{masters_triangle1}
I_1 &= I_{0, 0, 0, 0, 2, 2, 0}\,, \quad &I_2 &= I_{1, 0, 2, 0, 0, 2, 0}\,, \quad &I_3 &= I_{0, 2, 2, 0, 1, 0, 0}\,, \quad I_4 &= I_{0, 2, 1, 0, 2, 0, 0}\, , \\
I_5 &= I_{0, 2, 1, 1, 1, 0, 0}\,, \quad &I_6 &= I_{1, 0, 1, 0, 2, 1, 0}\,, \quad &I_7 &= I_{2, 0, 1, 0, 2, 1, 0}\,, \quad I_8 &= I_{0, 1, 1, 1, 1, 1, 0}\, , \\
I_9 &= I_{1, 0, 1, 1, 1, 1, 0}\,, \quad &I_{10} &= I_{1, 1, 1, 1, 1, 1, 0}\,, \quad &I_{11} &= I_{1, 2, 1, 1, 1, 1, 0} \, .   &
\end{gleichung}
The master integrals $\{ I_1, \hdots, I_9\}$ are  polylogarithmic and their differential equations can be brought into canonical form by standard methods. 
Explicitly, by studying the leading singularities of the integrals above, one can find the rotation 
\begin{gleichung}
    \textbf{T}^\text{sub} =\begin{pmatrix} 
           \epsilon ^2 & 0 & 0 & 0 & 0 & 0 & 0 & 0 & 0 \\
 0 & -\frac{\epsilon ^2}{z} & 0 & 0 & 0 & 0 & 0 & 0 & 0 \\
 0 & 0 & \frac{\sqrt{1+4 z} \epsilon ^2}{z} & \frac{\sqrt{1+4 z} \epsilon ^2}{2 z} & 0 & 0 & 0 & 0 & 0 \\
 0 & 0 & 0 & -\frac{\epsilon ^2}{z} & 0 & 0 & 0 & 0 & 0 \\
 0 & 0 & 0 & 0 & -\frac{\epsilon ^3}{z} & 0 & 0 & 0 & 0 \\
 0 & 0 & 0 & 0 & 0 & -\frac{\epsilon ^3}{z} & 0 & 0 & 0 \\
 \frac{\sqrt{1-4 z} \epsilon ^3}{2 z (1+2 \epsilon )} & 0 & 0 & 0 & 0 & 0 & \frac{\sqrt{1-4 z} \epsilon ^2}{z^2} & 0 & 0 \\
 0 & 0 & 0 & 0 & 0 & 0 & 0 & -\frac{\epsilon ^4}{z} & 0 \\
 0 & 0 & 0 & 0 & 0 & 0 & 0 & 0 & -\frac{\epsilon ^4}{z}
    \end{pmatrix}
\end{gleichung}
by which a complete $\eps$-factorisation of the subsectors can be obtained~\cite{vonManteuffel:2017hms}.

Now we can focus on the last two master integrals. As for the sunrise, integral $I_{10}$
is chosen by studying the leading singularities of the top sector integrals 
and selecting an integrand that can be put in the schematic form
of eq.~\eqref{eq:sunLSE4}. The candidate for $I_{11}$ is instead selected by adding a dot
on a massive propagator, which does not worsen its IR behaviour. Moreover, it is convenient to normalise the polynomial defining the underlying elliptic curve (c.f.~\eqref{eq:pure_psi0}) such that its highest monomial has coefficient one. This implies a normalisation of the master integrals $I_{10}$ and $I_{11}$ by $-1/z$.
Then we consider their homogeneous differential equations 
\begin{gleichung}
\label{eq:elliptriangle1}
    \frac{\mathrm d}{\mathrm dz} \begin{pmatrix} \tilde{\mathfrak I}_{10} \\ \tilde{\mathfrak I}_{11} \end{pmatrix} = \textbf{GM}^{\mathcal E} \begin{pmatrix} \tilde{\mathfrak I}_{10} \\ \tilde{\mathfrak I}_{11} \end{pmatrix} \quad\text{with}\quad \textbf{GM}^{\mathcal E} = 
    \begin{pmatrix}
         \frac{1}{z} & \frac{4}{z} \\
 \frac{1}{1-16 z} & \frac{1}{z (1-16 z)}
    \end{pmatrix} \, .
\end{gleichung}
This system is of elliptic type and has as fundamental solutions
\begin{gleichung}
\label{soltri1}
    \varpi_0(z) &=   \frac2\pi z \, \mathrm{K}(16z)        &&=   z(1+4 z+36 z^2+400z^3) + \mathcal O(z^5) \, ,\\
    \varpi_1(z) &=   -2z \, \mathrm{K}(1-16z)        &&= \varpi_0(z)\log(z) + 8 z^2+84 z^3+\frac{\num{2960}}{3}z^4 + \mathcal O(z^5) \, .
\end{gleichung}
From this we can consider the Wronskian matrix
\begin{gleichung}
\label{wronskiantri1}
    \textbf{W} = \begin{pmatrix}
        1 & 0 \\
 -\frac{1}{4} & \frac{z}{4}
    \end{pmatrix} \begin{pmatrix}
        \varpi_0 & \varpi_1 \\
        \varpi_0' & \varpi_1'
    \end{pmatrix}
\end{gleichung}
satisfying \eqref{eq:elliptriangle1}. We write its semi-simple part as
\begin{gleichung}
\label{semiwtri1}
    \textbf{W}^\text{ss} = \begin{pmatrix}
        1 & 0 \\
 -\frac{1}{4} & \frac{z}{4}
    \end{pmatrix}  \begin{pmatrix}
        \varpi_0 & 0 \\
        \varpi_0' & \frac{z}{(1-16 z)\varpi_0}
    \end{pmatrix} \, ,
\end{gleichung}
and as before we rotate our basis of masters by  the inverse of $\textbf{W}^\text{ss}$ followed by a suitable $\eps$-rescaling of the integrals. After that, one immediately sees that 
the term proportional to $\varpi_0'(z)$ and the non-$\eps$-factorised term can be removed by shifting the integrals by a total derivative. The full rotation can then be obtained as
\begin{gleichung}
\label{rottri1}
    \textbf{T} = 
    \begin{pmatrix}
        \mathbbm 1_{9\times 9} & 0 \\
        0 & \begin{matrix} 1 & 0 \\
                       \frac{1-24 z}{z^2}\varpi_0^2  & 1 \end{matrix}
    \end{pmatrix} 
    \begin{pmatrix}
        \mathbbm 1_{9\times 9} & 0 \\
         0     & \begin{matrix} \eps^4 & 0 \\
                       0  & \eps^3 \end{matrix}
    \end{pmatrix} 
    \begin{pmatrix}
        \mathbbm 1_{9\times 9} & 0 \\
         0                                & \textbf{W}^\text{ss}
    \end{pmatrix} ^{-1}
    \begin{pmatrix}
        \textbf{T}^\text{sub} & 0 \\
         0                                & -\frac1z\mathbbm 1_{2\times 2}
    \end{pmatrix}\,.
\end{gleichung}
With this, the $\eps$-factorised GM equations are given by
\begin{gleichung}
\label{epsgmtri1}
    \frac{\mathrm d}{\mathrm dz} \underline J &= \eps\, \textbf{GM}^{\eps}\, \underline J \quad\text{with}\quad \underline J = \mathbf{T}\, \underline I \, ,
\end{gleichung}
where the Gauss-Manin connection $\textbf{GM}^{\eps}$ is given in appendix \ref{app_triangle} eq. \eqref{gmepstri1}. We see here that in this single-scale problem 
we do not need additional new functions.
In fact, only the holomorphic solution $\varpi_0$ is needed, while its derivative  $\varpi_0'$ does \emph{not} appear in $\textbf{GM}^{\eps}$. The absence of the derivative
of the period is an indication that all differential forms are independent under integration
by parts identities. Nevertheless, we stress that we have not proven this last statement formally
and we cannot exclude that non-trivial relations exist.

\subsection{A second single-scale elliptic three-point function}
\label{ssec:triagnle2}

%-----Picture Triangle2 Graph-------------
\begin{figure}[!h]
\centering
\begin{tikzpicture}
\coordinate (links) at (-0.5,0);
\coordinate (mitte) at (0.5,0);
\coordinate (mmitte) at (1.92,-0.1); 
\coordinate (mmmitte) at (2.05,0.1); 
\coordinate (oben) at (1.5,1);
\coordinate (unten) at (1.5,-1);
\coordinate (obenr) at (2.5,1);
\coordinate (untenr) at (2.5,-1);
\coordinate (obenrr) at (3.5,1);
\coordinate (untenrr) at (3.5,-1);
\begin{scope}
\draw [-, dashed,postaction={decorate}] (oben) to [bend right=0]  (obenr);
\draw [-, dashed,postaction={decorate}] (untenr) to [bend right=0]  (unten);
\draw [-, thick,postaction={decorate}] (oben) to [bend right=0]  (untenr);
\draw [-, thick,postaction={decorate}] (obenr) to [bend right=0]  (mmmitte);
\draw [-, thick,postaction={decorate}] (unten) to [bend right=0]  (mmitte);
\draw [-, dashed,postaction={decorate}] (obenrr) to [bend right=0]  (obenr);
\draw [-, dashed,postaction={decorate}] (untenrr) to [bend right=0]  (untenr);
\draw [-, dashed,postaction={decorate}] (mitte) to [bend right=0]  (oben);
\draw [-, dashed,postaction={decorate}] (unten) to [bend right=0]  (mitte);
\end{scope}
\begin{scope}
\draw [-, line width=2pt,postaction={decorate}] (mitte) to [bend right=0]  (links);
\end{scope}
\node (d1) at (0.9,0.7) [font=\scriptsize, text width=.2 cm]{1};
\node (d2) at (0.9,-0.7) [font=\scriptsize, text width=.2 cm]{3};
\node (d3) at (2,1.2) [font=\scriptsize, text width=.2 cm]{2};
\node (d4) at (2,-1.2) [font=\scriptsize, text width=.2 cm]{4};
\node (d5) at (2,1.2) [font=\scriptsize, text width=.2 cm]{2};
\node (d6) at (1.6,0.4) [font=\scriptsize, text width=.2 cm]{5};
\node (d7) at (2.5,0.4) [font=\scriptsize, text width=.2 cm]{6};
\node (d8) at (0.05,0.2) [font=\small, text width=.2 cm]{$s$};
\node (d9) at (3,1.3) [font=\small, text width=.2 cm]{$p_1^2$};
\node (d10) at (3,-1.3) [font=\small, text width=.2 cm]{$p_2^2$};
\end{tikzpicture}
\caption{The graph of non-planar triangle number 2.}
\label{fig:triangle2}
\end{figure}
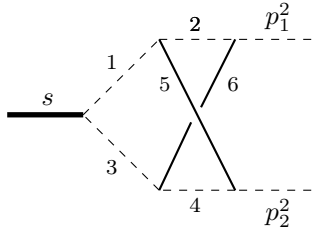
%------------------------

\noindent
The next family of Feynman integrals we consider is given by another triangle graph (see figure \ref{fig:triangle2}) but with a different mass configuration of the propagators and numerator
\begin{gleichung}
\label{def_triangle2family}
D_1 &= (k_1-p_1)^2\,, \quad &D_2 &= (k_2-p_1)^2\,, \quad  &D_3 &= (k_1+p_2)^2\,, \\
D_4 &= (k_1-k_2+p_2)^2\,, \quad &D_5 &= (k_1-k_2)^2-m^2\,, \quad  &D_6 &= k_2^2-m^2\, , \\
N_1 &= k_1^2 \, .
\end{gleichung}
For the external momenta, we again consider $p_1^2=p_2^2=0$ and work with the dimensionless variable 
\begin{equation}
z=\frac{s}{m^2} = \frac{(p_1+p_2)^2}{m^2} \, .
\end{equation}
As before, we study these integrals in $d=4-2\eps$ and set for simplicity $m=1$. We recall here that this class of integrals contributes to the so-called electroweak 
form factor and was first studied in~\cite{Aglietti:2007as} and later on in~\cite{Broedel:2019hyg}.\footnote{Also in this case, we thank Xing Wang for having shared his results for this integral family with us prior to publication~\cite{Xing:2023}.}

Integration by part identities show that there are $18$ independent master integrals $\underline I=\{ I_1, \hdots, I_{18}\}$ for which we choose the following starting basis
\begin{gleichung}
\label{masters_triangle2}
I_1 &= I_{0, 0, 0, 0, 2, 2, 0}\,, \ \ &I_2 &= I_{0, 0, 2, 1, 0, 2, 0}\,, \ \ &I_3 &= I_{2, 0, 1, 0, 0, 2, 0}\,, \ \ &I_4 &= I_{0, 2, 1, 0, 2, 0, 0}\, , \\
I_5 &= I_{0, 2, 2, 0, 1, 0, 0}\,, \ \ &I_6 &= I_{0, 1, 1, 0, 2, 1, 0}\,, \ \ &I_7 &= I_{1, 0, 1, 0, 2, 1, 0}\,, \ \ &I_8 &= I_{1, 1, 1, 2, 0, 0, 0}\, , \\
I_9 &= I_{0, 1, 2, 0, 1, 1, 0}\,, \ \ &I_{10} &= I_{2, 0, 1, 0, 2, 1, 0}\,, \ \ &I_{11} &= I_{0, 1, 1, 1, 1, 1, 0}\,, \ \ &I_{12} &= I_{1, 0, 1, 1, 1, 1, 0}\,, \\
I_{13} &= I_{1, 1, 1, 1, 0, 1, 0}\,, \ \ &I_{14} &= I_{0, 2, 1, 1, 1, 1, 0}\,, \ \ &I_{15} &= I_{2, 1, 1, 1, 0, 1, 0}\,, \ \ &I_{16} &= I_{1, 1, 1, 1, 1, 1, 0} \,,  \\
I_{17} &= I_{1, 1, 1, 1, 2, 1, 0}\,, \ \ &I_{18} &= I_{1, 1, 1, 1, 1, 1, -1} \, . && 
\end{gleichung}
The first fifteen integrals $\{I_1, \hdots, I_{15}\}$ are of polylogarithmic type and a rotation $\textbf{T}^\text{sub}$ that puts them in canonical form is readily found by standard methods. The explicit form of this transformation is given in appendix \ref{app_triangle}, see eq. \eqref{rottri2}.

Let us focus instead on the elliptic top sector, which this time contains three master integrals $I_{16}, I_{17}, I_{18}$. A leading singularity analysis for these integrals also suggests to normalise them with a factor of $z$, such that the quartic polynomial defining the underlying elliptic curve has the highest monomial normalised to one. This also has the effect of decoupling completely the integral $I_{18}$ in $d=4$ from the other two, since the numerator that defines it,
generates a residue at infinity exactly equal to $1/z$.
With this, the maximal cuts of the normalised integrals $\tilde I_{16}, \tilde I_{17}$ satisfy an elliptic homogeneous set of differential equations
\begin{gleichung}
\label{eq:elliptriangle2}
    \frac{\mathrm d}{\mathrm dz} \begin{pmatrix} \tilde {\mathfrak I}_{16} \\ \tilde {\mathfrak I}_{17} \end{pmatrix} = \textbf{GM}^{\mathcal E} \begin{pmatrix} \tilde {\mathfrak I}_{16} \\ \tilde {\mathfrak I}_{17} \end{pmatrix} \quad\text{with}\quad \textbf{GM}^{\mathcal E} = 
    \begin{pmatrix}
         -\frac{1}{z} & -\frac{2}{z} \\
 \frac{4-z}{(1-z) (8+z) z} & \frac{8+z^2}{(1-z) (8+z) z}
    \end{pmatrix} \, .
\end{gleichung}
The two fundamental solutions are given by\footnote{We have neglected terms proportional to $\log(2)$ in the expansion of the second elliptic function which is a result of the variable $z$ requiring a rescaling to yield a non-integer coefficient expansion of the first elliptic function. As it was observed in \cite{Aglietti:2007as} one can also make a change of variables and a subsequent rotation to relate the two elliptic functions \eqref{soltri2} of this triangle family to the elliptic functions of the equal-mass sunrise graph in two dimensions.}
\begin{gleichung}
\label{soltri2}
    \varpi_0(z) &=   \frac{16}\pi  \frac{\mathrm{K}\left(16 \frac{z^{3/2} \sqrt{8+z}}{8 \left(8-z \left(4+z-\sqrt{z} \sqrt{8+z}\right)\right)}\right)}{\sqrt{8 \left(8-z \left(4+z-\sqrt{z} \sqrt{8+z}\right)\right)}}        &&=   1+\frac{z}{4}+\frac{5 z^2}{32} + \mathcal O(z^3) \\
    \varpi_1(z) &=   -\frac{32}{3}\frac{\mathrm{K}\left(1-16 \frac{z^{3/2} \sqrt{8+z}}{8 \left(8-z \left(4+z-\sqrt{z} \sqrt{8+z}\right)\right)}\right)}{\sqrt{8 \left(8-z \left(4+z-\sqrt{z} \sqrt{8+z}\right)\right)}}         &&= \varpi_0(z)\log(z) + \frac{3 z}{8}+\frac{33 z^2}{128} +  \mathcal O(z^3) \, .
\end{gleichung}
The semi-simple part of the corresponding Wronskian is given by
\begin{gleichung}
\label{semiwtri2}
    \textbf{W}^\text{ss} = \begin{pmatrix}
        1 & 0 \\
 -\frac{1}{2} & -\frac{z}{2}
    \end{pmatrix}  \begin{pmatrix}
        \varpi_0 & 0 \\
        \varpi_0' & \frac{8}{(1-z) (8+z) z\varpi_0}
    \end{pmatrix} \, .
\end{gleichung}
From this we can construct, by the same reasoning as before, the full rotation matrix
\begin{gleichung}
    \textbf{T} &= 
    \begin{pmatrix}
        \mathbbm 1_{15\times 15} & 0\phantom{XX} \\
         0     & \begin{matrix} 1 & 0 & 0 \\
                       -\frac{2}{3}(1-z)\varpi_0  & 1 & 0 \\
                       -\frac{1}{24} \left(76-44 z-5 z^2\right)\varpi_0^2 & 0 & 1\end{matrix}
    \end{pmatrix} 
    \begin{pmatrix}
        \mathbbm 1_{15\times 15} & 0 \\
         0     & \begin{matrix} \eps^4 & 0 & 0 \\
                       0  & 0 & \eps^4 \\
                       0  & \eps^3  & 0\end{matrix}
    \end{pmatrix} \\[1ex]
    &  \hspace{1cm}\times
    \begin{pmatrix}
        \mathbbm 1_{15\times 15} & 0 \\
         0                                & \textbf{W}^\text{ss}
    \end{pmatrix} ^{-1}
    \begin{pmatrix}
        \textbf{T}^\text{sub} & 0 \\
         0                                & z\mathbbm 1_{3\times 3}
    \end{pmatrix} \, .
\end{gleichung}
The final $\eps$-factorised GM differential equation is then given by 
\begin{gleichung}
\label{epsgmtri2}
    \frac{\mathrm d}{\mathrm dz} \underline J &= \eps\, \textbf{GM}^{\eps}\, \underline J \quad\text{with}\quad \underline J = \mathbf{T}\, \underline I \, ,
\end{gleichung}
where the GM connection $\textbf{GM}^{\eps}$ can be found in appendix \ref{app_triangle} eq. \eqref{gmepstri2}. We stress that once again, $\varpi_0'$ does \emph{not} appear in $\textbf{GM}^{\eps}$, an important prerequisite to conjecture that all differential forms in eq.~\eqref{epsgmtri2} are linearly independent on the field of algebraic functions. 

As for the equal-mass sunrise and the first triangle family studied in the previous subsection, at every step of the rotation to an $\eps$-factorised basis we need only $\varpi_0$ and $\varpi_0'$ and no extra function has to be introduced. This might be surprising since, contrary to the two previous cases, there is a third master integral in the top sector, whose integrand on the maximal cut at $\eps=0$ corresponds to a differential of the third kind on an elliptic curve. However, upon explicit integration over the cycles of the elliptic curve, the resulting elliptic integral of the third kind might degenerate to linear combinations of $\varpi_0$, $\varpi_1$ and their derivatives, which would explain why no new function is required to achieve the factorisation of $\eps$. For the region $0<z<1$, we verified that upon integration over the cycle of the elliptic curve for which $\text{MaxCut}(\tilde{I}_{16}) \propto \varpi_0$, one finds indeed such a relation:
\begin{equation}
    \text{MaxCut}(\tilde{I}_{18}) = \frac{2}{3}(1-z) \, \text{MaxCut}(\tilde{I}_{16}) + c \, ,
\end{equation}
where $c$ is just a constant.
 
\subsection{Two-parameter triangle}
\label{ssec:triagnle12par}
We have seen that in the different mass sunrise graph, which was also the only multi-parameter problem analysed so far, new objects had to be introduced to achieve a complete
$\eps$-factorisation of the differential equations.
In this and the next subsection, we study other multi-parameter problems, related to more complicated processes, to see if the same pattern can be observed.
As a two-parameter family, we consider again the non-planar triangle from figure \ref{fig:triangle1}, but this time 
with one more external leg off-shell, i.e. $p_1^2=0$ and $p_2^2=M^2$. The problem depends now on two dimensionless parameters, 
which we choose to be $z_1=s/m^2$ and $z_2=M^2/m^2$.

The family admits in this case $22$ master integrals and a suitable starting basis $\underline I=\{ I_1,\hdots, I_{22} \}$ is given by
\begin{gleichung}
\label{masters_triangle12par}
I_1 &= I_{0, 2, 0, 2, 0, 0, 0}\,, \ \ &I_2 &= I_{2, 2, 1, 0, 0, 0, 0}\,, \ \ &I_3 &= I_{0, 2, 0, 1, 2, 0, 0}\,, \ \ I_4 &= I_{0, 2, 1, 0, 2, 0, 0}\, , \\
I_5 &= I_{0, 1, 2, 0, 2, 0, 0}\,, \ \ &I_6 &= I_{0, 0, 1, 0, 2, 2, 0}\,, \ \ &I_7 &= I_{0, 0, 2, 0, 2, 1, 0}\,, \ \ I_8 &= I_{1, 0, 2, 2, 1, 0, 0}\, , \\
I_9 &= I_{1, 1, 1, 2, 0, 0, 0}\,, \ \ &I_{10} &= I_{2, 1, 1, 2, 0, 0, 0}\,, \ \ &I_{11} &= I_{0, 1, 1, 0, 2, 1, 0}\,, \ \ I_{12} &= I_{0, 2, 1, 1, 1, 0, 0}\, , \\
I_{13} &= I_{0, 3, 1, 1, 1, 0, 0}\,, \ \ &I_{14} &= I_{0, 2, 1, 1, 2, 0, 0}\,, \ \ &I_{15} &= I_{1, 1, 1, 1, 0, 1, 0}\,, \ \ I_{16} &= I_{1, 1, 1, 1, 1, 0, 0}\, , \\
I_{17} &= I_{1, 1, 1, 2, 1, 0, 0}\,, \ \ &I_{18} &= I_{1, 2, 1, 1, 1, 0, 0}\,, \ \ &I_{19} &= I_{0, 1, 1, 1, 1, 1, 0}\,, \ \ I_{20} &= I_{0, 1, 1, 1, 2, 1, 0}\, , \\
I_{21} &= I_{1, 1, 1, 1, 1, 1, 0}\,, \ \ &I_{22} &= I_{1, 1, 1, 2, 1, 1, 0} \, .
\end{gleichung}
The master integrals $\{ I_1, \hdots, I_{20}\}$ are of polylogarithmic nature and their differential equations can be brought into canonical form by standard methods. The rotation matrix which brings the subsectors into canonical form can be found in an ancillary file to this manuscript. 

Again, the interesting step consists in the $\eps$-factorisation of the top sector 
containing the two integrals $\{ I_{21}, I_{22}\}$. Their maximal cuts satisfy the following homogeneous equations
\begin{gleichung}
\label{gmtri22par}
    \mathrm d \begin{pmatrix} \mathfrak I_{21} \\ \mathfrak I_{22} \end{pmatrix} &= \left(  \mathbf{GM}^{\mathcal E, z_1}\, \mathrm dz_1 + \mathbf{GM}^{\mathcal E, z_2}\, \mathrm dz_2 \right)\begin{pmatrix} \mathfrak I_{21} \\ \mathfrak I_{22} \end{pmatrix} \quad\text{with}   \\
\mathbf{GM}^{\mathcal E, z_1} &= \begin{pmatrix} -\frac{2}{z_1-z_2} & -\frac{4 \left(z_1+z_2\right)}{\left(z_1-z_2\right){}^2} \\
 \frac{z_1+z_2}{z_1 \Delta(\underline z) } & -\frac{2 z_1^3-4 z_1 \left(8-z_2\right) z_2-z_2^3+z_1^2 \left(16-5 z_2\right)}{z_1 \left(z_1-z_2\right) \Delta(\underline z) } \end{pmatrix}, \\
   \mathbf{GM}^{\mathcal E, z_2} &= \begin{pmatrix} \frac{2}{z_1-z_2} & \frac{8 z_1}{\left(z_1-z_2\right){}^2} \\
 -\frac{2}{\Delta(\underline z) } & \frac{z_1^2+z_2^2-2 z_1 \left(8+z_2\right)}{\left(z_1-z_2\right) \Delta(\underline z) }\end{pmatrix} \, , \\
    \Delta(\underline z) &=  z_1^2+2 z_1 \left(8-z_2\right)+z_2^2  \, .
\end{gleichung}
We can now construct the rotation matrix which brings this $(2\times2)$-system into $\eps$-factorised form
\begin{gleichung}
    \textbf{T}^{2\times 2} &= \begin{pmatrix} 1 & 0 \\
 \frac{z_1^3+z_1^2 \left(8-3 z_2\right)+3 z_1 \left(-8+z_2\right) z_2-z_2^3}{z_1+z_2}\varpi_0^2 & 1 \end{pmatrix} 
 \begin{pmatrix}   \eps & 0 \\ 0 & 1  \end{pmatrix} 
 \begin{pmatrix}
     \varpi_0 & 0 \\ \partial_1\varpi_0 & \frac{z_1+z_2}{z_1 \left(z_1-z_2\right)\Delta(\underline z)\varpi_0}
 \end{pmatrix}^{-1} \\
 &\hspace{3cm} \times \begin{pmatrix}
     1 & 0 \\
 -\frac{1+2 \epsilon }{z_1-z_2} & -\frac{4 \left(z_1+z_2\right)}{\left(z_1-z_2\right){}^2}
 \end{pmatrix}  \begin{pmatrix} z_1 -z_2 & 0 \\ 0 & z_1-z_2 \end{pmatrix} \, .
\end{gleichung}
To build this rotation we followed the usual steps: First, we have normalised by a factor of $z_1-z_2$ such that again the highest monomial of the quartic polynomial defining the underlying elliptic curve is normalised to one. Afterwards, we perform a rotation to go to the derivative basis with respect to $z_1$ in the top sector. We then multiplied by the inverse semi-simple part of the resulting Wronskian, which contains as the holomorphic fundamental solution the elliptic function
\begin{gleichung}
    \varpi_0(z_1,z_2) = \frac{\mathrm{K}\left(  \frac{\left(z_1+\sqrt{16 z_1+\left(z_1-z_2\right){}^2}-z_2\right){}^2}{4 \sqrt{16 z_1+\left(z_1-z_2\right){}^2}
   \left(z_1-z_2\right)} \right)}{\sqrt{\sqrt{16 z_1+\left(z_1-z_2\right){}^2} \left(z_1-z_2\right)}} \, .
\end{gleichung}
As before, we hide the dependence on the variables $z_1,z_2$ in the periods and abbreviate $\varpi_0=\varpi_0(z_1,z_2)$ whenever convenient.
Finally, we removed the terms which are not proportional to $\eps$ by simple rotations.

To achieve a complete $\eps$-factorised form for the full Gauss-Manin system, we have to perform additional rotations in the subsectors 
contributing to the inhomogeneous part of the differential equation for the top sector. 
Compared to the single-scale problem analysed before, in this case, not all non-$\eps$-factorised terms can be removed by simple rotations that do not involve any new function. There are in fact two terms which can only be removed by introducing a new object, say $G\coloneqq G(z_1,z_2)$, defined by the two partial derivatives
\begin{align}
\label{eq:deqsGz1z2} 
    \partial_{z_1} G &= \frac{ 4\left(z_1-z_2\right) \sqrt{z_2-4} \sqrt{z_2}}{\left(z_1+z_2\right){}^2}\, \varpi_0 \,, \\
    \partial_{z_2} G &= \frac{2 \left(z_1 \left(16+z_1-3 z_2\right) \left(z_1-z_2\right)\right)}{\left(z_1+z_2\right){}^2 \sqrt{z_2-4} \sqrt{z_2}} \, \varpi_0 + \frac{2 z_1 \left(16 z_1+\left(z_1-z_2\right){}^2\right) \left(z_1-z_2\right)}{\left(z_1+z_2\right){}^2 \sqrt{z_2-4} \sqrt{z_2}}\, \partial_1\varpi_0 \, . \nonumber   
\end{align} 
One can easily see that the first equation is solved by
\begin{equation}
    G(z_1,z_2) = 4\sqrt{z_2-4} \sqrt{z_2}\int_0^{z_1} \frac{ \left(u-z_2\right) }{\left(u+z_2\right){}^2}\, \varpi_0(u,z_2) \, \mathrm du \, .
\end{equation}
The second equation in \eqref{eq:deqsGz1z2} is then satisfied by using the differential equations for $\varpi_0$. 

We stress here that, at variance with the two-parameter sunrise case, where a new function $G$ was required for the $\eps$-factorisation of the $(3\times3)$-homogeneous system,
in this case
we do not need a new function to obtain an $\eps$-factorised form for the homogeneous system associated with the top sector. 
Instead, the new function originates from the subtopologies, which can be understood as follows:
First, the additional scale that this problem depends on does not increase the number of master integrals in the top sector. 
This can be interpreted as the fact that,
contrary to the sunrise, there is no extra residue in the integrand of the maximal cut, which is linearly independent under integration by parts. This implies that there is no contribution from an elliptic integral of the third kind to the homogeneous differential equations.
Second, there is instead a different singularity structure in the subsectors, namely a new pole 
compared to the ones found in the homogeneous differential equations for the top sector.  
Integrating over this pole gives a contribution formally similar to an extra residue in the homogeneous equation.
The full rotation matrix and the final form of the GM matrices can be found in an ancillary file.
\vspace{3ex}

\subsection{Three-parameter double box}
\label{ssec:doubbox}

%-----Picture Double Box Graph-------------
\begin{figure}[!h]
\centering
\begin{tikzpicture}
\coordinate (llo) at (-2.5,1);
\coordinate (lo) at (-1.5,1);
\coordinate (mo) at (0,1);
\coordinate (ro) at (1.5,1);
\coordinate (rro) at (2.5,1);
\coordinate (llu) at (-2.5,-1);
\coordinate (lu) at (-1.5,-1);
\coordinate (mu) at (0,-1);
\coordinate (ru) at (1.5,-1);
\coordinate (rru) at (2.5,-1);
\begin{scope}
\draw [-, dashed,postaction={decorate}] (llo) to [bend right=0]  (lo);
\draw [-, dashed,postaction={decorate}] (rro) to [bend right=0]  (ro);
\draw [-, dashed,postaction={decorate}] (llu) to [bend right=0]  (lu);
\draw [-, line width=2pt,postaction={decorate}] (rru) to [bend right=0]  (ru);
\draw [-, dashed,postaction={decorate}] (mo) to [bend right=0]  (mu);
\draw [-, thick,postaction={decorate}] (lo) to [bend right=0]  (ro);
\draw [-, thick,postaction={decorate}] (lu) to [bend right=0]  (ru);
\draw [-, thick,postaction={decorate}] (lo) to [bend right=0]  (lu);
\draw [-, thick,postaction={decorate}] (ro) to [bend right=0]  (ru);
\end{scope}
\node (d1) at (-0.75,1.2) [font=\scriptsize, text width=.2 cm]{1};
\node (d2) at (0.75,1.2) [font=\scriptsize, text width=.2 cm]{3};
\node (d3) at (-0.75,-1.2) [font=\scriptsize, text width=.2 cm]{2};
\node (d4) at (0.75,-1.2) [font=\scriptsize, text width=.2 cm]{4};
\node (d5) at (-1.7,0) [font=\scriptsize, text width=.2 cm]{5};
\node (d6) at (-0.2,0) [font=\scriptsize, text width=.2 cm]{6};
\node (d7) at (1.3,0) [font=\scriptsize, text width=.2 cm]{7};
\node (d8) at (-2,1.3) [font=\small, text width=.2 cm]{$p_1^2$};
\node (d9) at (-2,-1.3) [font=\small, text width=.2 cm]{$p_2^2$};
\node (d10) at (2,1.3) [font=\small, text width=.2 cm]{$p_3^2$};
\node (d11) at (2,-1.3) [font=\small, text width=.2 cm]{$p_4^2$};
\end{tikzpicture}
\caption{The double box graph.}
\label{fig:doublebox}
\end{figure}
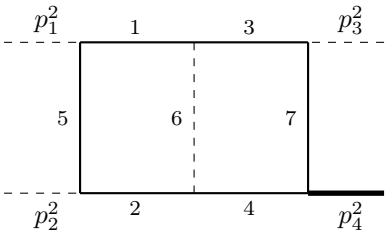
%------------------------

\noindent
As a final example, we consider the double box graph depicted in figure \ref{fig:doublebox}. The associated integral family is defined by the following propagators and irreducible numerators:
\begin{equation}
\label{def_doubleboxfamily}
\begin{aligned}
D_1 &= k_1^2-m^2\,, \quad &D_2 &= (k_1+p_1+p_2)^2-m^2\,, \quad  &D_3 &= k_2^2-m^2\,,  \\
D_4 &= (k_2+p_1+p_2)^2-m^2\,, \quad &D_5 &= (k_1+p_1)^2-m^2\,, \quad  &D_6 &= (k_1-k_2)^2\,, \\
D_7 &= (k_2-p_3)^2-m^2\, , \quad &N_1 &= (k_2+p_1)^2\, , \quad &N_2 &= (k_1-p_3)^2 \, .
\end{aligned}
\end{equation}
The external momenta satisfy $p_1^2=p_2^2=p_3^2=0$ and $p_4^2=M^2$.

With Mandelstam variables defined by $s=(p_1+p_2)^2$ and $t=(p_1+p_3)^2$, the family depends on a total of four dimensionful or, by dimensional analysis, three dimensionless variables. Further, we set $d=4-2\eps$ for this example.

This graph contributes to the planar corrections to the production of a Higgs boson and a jet through a loop of massive quarks
and was studied at length in the literature, see~\cite{Bonciani_2016,Frellesvig:2019byn}, where a complete calculation could only be achieved numerically.\footnote{Note that compared to~\cite{Bonciani_2016}, our propagator definitions~\eqref{def_doubleboxfamily} differ by a sign.} 
In the literature, a total of 73 master integrals had been identified. 
A new reduction performed with \textsc{Kira 2.0}~\cite{Klappert_2021} reveals an additional 
relation, originating from a higher sector, which reduces this number to 72. The relation reads 
\begin{equation}
\label{addrelation}
    I_{0,1,0,0,1,1,2,0,0} =\frac{s}{M^2-t} I_{0,1,2,0,1,1,0,0,0}+\frac{M^2-s}{M^2-t} I_{0,2,1,0,0,1,1,0,0} - \frac{t}{M^2-t} I_{1,0,0,0,1,1,2,0,0} \, .
\end{equation}

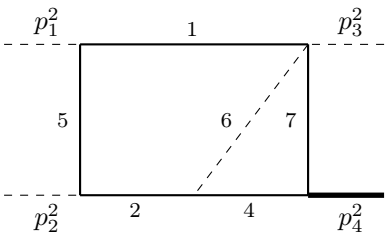
\begin{figure}[!h]
\centering
\begin{tikzpicture}
\coordinate (llo) at (-2.5,1);
\coordinate (lo) at (-1.5,1);
\coordinate (mo) at (0,1);
\coordinate (ro) at (1.5,1);
\coordinate (rro) at (2.5,1);
\coordinate (llu) at (-2.5,-1);
\coordinate (lu) at (-1.5,-1);
\coordinate (mu) at (0,-1);
\coordinate (ru) at (1.5,-1);
\coordinate (rru) at (2.5,-1);
\begin{scope}
\draw [-, dashed,postaction={decorate}] (llo) to [bend right=0]  (lo);
\draw [-, dashed,postaction={decorate}] (rro) to [bend right=0]  (ro);
\draw [-, dashed,postaction={decorate}] (llu) to [bend right=0]  (lu);
\draw [-, line width=2pt,postaction={decorate}] (rru) to [bend right=0]  (ru);
\draw [-, dashed,postaction={decorate}] (ro) to [bend right=0]  (mu);
\draw [-, thick,postaction={decorate}] (lo) to [bend right=0]  (ro);
\draw [-, thick,postaction={decorate}] (lu) to [bend right=0]  (ru);
\draw [-, thick,postaction={decorate}] (lo) to [bend right=0]  (lu);
\draw [-, thick,postaction={decorate}] (ro) to [bend right=0]  (ru);
\end{scope}
\node (d1) at (0,1.2) [font=\scriptsize, text width=.2 cm]{1};
\node (d3) at (-0.75,-1.2) [font=\scriptsize, text width=.2 cm]{2};
\node (d4) at (0.75,-1.2) [font=\scriptsize, text width=.2 cm]{4};
\node (d5) at (-1.7,0) [font=\scriptsize, text width=.2 cm]{5};
\node (d6) at (0.45,0) [font=\scriptsize, text width=.2 cm]{6};
\node (d7) at (1.3,0) [font=\scriptsize, text width=.2 cm]{7};
\node (d8) at (-2,1.3) [font=\small, text width=.2 cm]{$p_1^2$};
\node (d9) at (-2,-1.3) [font=\small, text width=.2 cm]{$p_2^2$};
\node (d10) at (2,1.3) [font=\small, text width=.2 cm]{$p_3^2$};
\node (d11) at (2,-1.3) [font=\small, text width=.2 cm]{$p_4^2$};
\end{tikzpicture}
\caption{Feynman graph of the elliptic sector inside the double box.}
\label{fig:elldoubbox}
\end{figure}

In~\cite{Bonciani_2016}, 64 independent candidate integrals for a canonical basis for all but one subsector 
had been identified (in the notation of~\cite{Bonciani_2016}, $f^A_{19}$ is no longer independent and we remove it from our basis). 
Moreover, four suitable candidates on the maximal cut of the top topology have been derived. 
The remaining four integrals lie in an elliptic subsector depicted in figure~\ref{fig:elldoubbox}, which is obtained from the top topology by pinching line number three. Here, we wish to complement these results by applying our method to find suitable basis integrals 
for the elliptic sector, such that a complete basis satisfying $\eps$-factorised differential equations 
is at hand. In the following, we will give an overview of the relevant details of each step.

%-----Picture Elliptic Sector in Double Box Graph-------------
%------------------------

As a starting basis to apply our method, we use a slight variation of the integrals proposed in~\cite{Bonciani_2016,Primo:2016ebd}
\begin{equation}
\begin{aligned}
    I_{65} &= \eps^4 \ I_{1,1,0,1,1,1,1,0,0}\, ,\\
    I_{66} &= \eps^4 \ I_{2,1,0,1,1,1,1,0,0}\, ,\\
    I_{67} &= \eps^3 \ I_{1,1,0,2,1,1,1,0,0}\, ,\\
    I_{68} &= \eps^4 \ s \ I_{1,1,0,1,1,1,1,0,-1}\, .\\
\end{aligned}
\end{equation}
Their differential equations have the properties we desire. Concretely, at $\eps=0$, the differential equations for $I_{67}$ and $I_{68}$ only couple to $I_{65}$ and $I_{66}$, but not to each other. Moreover, the differential equations for the first two master integrals $I_{65}$ and $I_{66}$ do not contain contributions from the last two. Consequently, their $( 2 \times 2)$-system encodes the elliptic part of the sector. The particular choice of integrals is motivated by the properties of the maximal cuts of this basis in a loop-by-loop Baikov approach, which yields a one-fold integral representation;
\begin{itemize}
    \item The maximal cut of $I_{68}$ exhibits a non-vanishing constant residue at infinity, in contrast to the other three master integrals.
    \item The maximal cut of $I_{1,1,0,2,1,1,1,0,0}$ is proportional to $\eps$ which means that it vanishes in strictly $d=4$ space-time dimensions. Therefore, its differential equation at $\eps=0$ should not couple to $I_{1,1,0,1,1,1,1,0,0}, I_{2,1,0,1,1,1,1,0,0}$ and $I_{1,1,0,1,1,1,1,0,-1}$, as this would imply a non-trivial relation among their maximal cuts. However, this also means that rescaling it by a factor $1/ \eps$ with respect to the other three master integrals makes it a potential candidate for our starting basis and we can verify from the explicit form of the differential equations that it is indeed suitable. Further, the maximal cut of $I_{1,1,0,2,1,1,1,0,0}$ possesses a non-vanishing residue, in contrast to the other three master integrals. From this one can read out the normalisation factor
    \begin{equation}
    \label{eq:4ptnormsqrt}
    \frac{m^2}{2 m^2-M^2} \sqrt{s \, t \, M^2 \left(M^2-4 m^2\right) \left(s t + 4 m^2 (M^2-s-t)\right)} \, ,
    \end{equation}
    which has to be added to the transformation to achieve full $\eps$-factorisation. 
    \item The maximal cut of the first integral $I_{65}$ at $\eps=0$ can be written  in a form analogous to eq.~\eqref{eq:sunLSE4}, but where the elliptic polylogarithmic kernel  does not appear in the last integration but in the next to last integration. As discussed above, conjecturally this is still a good candidate for one of the master integrals. 
    \item Finally, the fourth integral $I_{66}$ was chosen with a dot on a massive propagator such that it is independent of the other three. This excluded putting the dot on the seventh propagator.
\end{itemize}
All subsequent steps to achieve the factorisation of $\eps$ proceed as in previous cases, without any additional complications. 
The elliptic function introduced by the rotation with the inverse of the semi-simple part of the Wronskian solution matrix 
can be chosen as
\begin{gleichung}
    \varpi_0 (s,t,M^2,m^2) &= \dfrac{\mathrm{K}\left(\dfrac{16 m^2 \sqrt{M^2 s t \left(s+t-M^2\right)}}{Z(s,t,M^2,m^2)}\right)}{\sqrt{s \ Z(s,t,M^2,m^2)}} \quad\text{with} \\
    Z(s,t,M^2,m^2) &= M^4 s+8 m^2 \sqrt{M^2 s t \left(s+t-M^2\right)}- \\
    &- 2 M^2 \left(2 m^2 (s-t)+s t\right)+t \left(s t-4 m^2 (s+t)\right)\, .
\end{gleichung}

Following the discussion in the previous section, in this case, as expected from the number of independent master integrals in the top sector, the $\eps$-factorisation of the homogeneous system requires the introduction of two new functions $G_1, G_2$. In addition, two new functions $G_3, G_4$ are required to factorise $\eps$ in the whole system, including the subtopologies. These can be written as
\begin{equation}
    G_i(s,t,M^2,m^2) =\int_0^{m^2} r_i(s,t,M^2,u) \ \varpi_0(s,t,M^2,u) \, \mathrm du 
\end{equation}
with 
\begin{equation}
\begin{aligned}
    r_1(s,t,M^2,m^2) &= -(s+t) \, ,\\
    r_2(s,t,M^2,m^2) &= \, \frac{M^4 s t^2 \ P_2(s,t,M^2,m^2)}{\left[s \, t \, M^2 \left(M^2-4 m^2\right) \left(s t + 4 m^2 (M^2-s-t)\right)\right]^{3/2}} \, ,\\ 
    P_2(s,t,M^2,m^2) &= \left(M^2-s\right) s \left(64 m^6+ \left(48 s-64 M^2\right)m^4 +(s-4 m^2) M^4\right)+  \\
    &\quad+ \left(64 m^6 \left(M^2-2 s\right)+s M^2 \left(8 m^2 M^2-4(M^2+5m^2)
    s+5s^2\right) \right)t -  \\
    &\quad-16 m^4 t\left(2 M^4-7 M^2 s+2 s^2\right)+ \left(2 M^2-4 m^2-s\right) \left(s-4 m^2\right)^2 t^2 \, ,\\
    r_3(s,t,M^2,m^2) &= \, \frac{s\ P_3(s,t,M^2,m^2)}{\left[s \left(m^4 s+2 m^2 (2 M^2-s-2 t) t+s t^2\right)\right]^{3/2}} \, ,\\ 
    P_3(s,t,M^2,m^2) &= m^6 s (s+t) (3 s+4 t)-s^2 t^2 \left(M^4-4 M^2 t+t (3 s+2 t)\right)+ \\
    &\quad+m^2 s t \left(-M^4 (s-4 t)-4 M^2 t (s+2 t)+t (s+2t) (9 s+8 t)\right)+ \\
    &\quad+ m^4 t \left(4 M^2 (s+2 t) (5 s+4 t)-(s+t) \left(9 s^2+40 s t+32 t^2\right)\right) \, ,\\
    r_4(s,t,M^2,m^2) &= \frac{2 (M^2-s) s^3}{\left[s (s-4 m^2)\right]^{3/2}} \, .\\
\end{aligned}
\end{equation}
The square-root in the denominator of $r_2(s,t,M^2,m^2)$ stems from the normalisation factor~\eqref{eq:4ptnormsqrt} required for the integral $I_{67}$ in our starting basis. The expressions for  $r_3(s,t,M^2,m^2)$ and $r_4(s,t,M^2,m^2)$ contain additional square-roots, generated by other sectors. Thereby, we would like to point out that the function $G_4$ is required to factorise $\eps$ in the contributions from the elliptic sector to the differential equations of the top sector.

Also in this case, as for all previous examples, the final $\eps$-factorised Gauss-Manin connection matrix does not contain any explicit dependence on derivatives of $\varpi_0$. The explicit rotations and expressions for the Gauss-Manin connection matrices are included in an ancillary file. 

%% file: Sec4_BeyondEll.tex
% !TEX encoding = UTF-8 Unicode

\section{Cases beyond a single elliptic curve}
\label{sec:beyondell}

In the previous sections, we have provided examples of how our procedure can be applied 
to multi-scale
elliptic problems to obtain fully $\epsilon$-factorised systems of differential equations.
Here, we want to test the applicability of our procedure beyond the elliptic case, considering families of Feynman graphs with different underlying geometries. 
We start with a family with two elliptic curves and then consider a three-loop example characterised by a K3 surface.

\subsection{The three-loop ice cone}
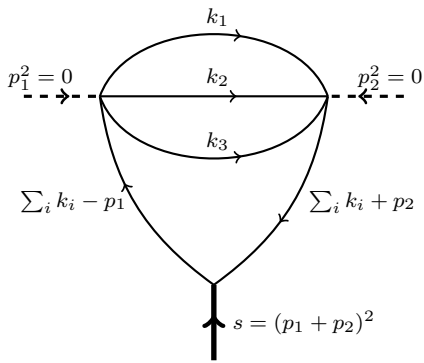
\begin{figure}[!h]
\centering
\centering
\begin{tikzpicture}
\coordinate (one) at (-2.5,0);
\coordinate (two) at (2.5,0);
\coordinate (One) at (-1.5,0);
\coordinate (Two) at (1.5,0);
\coordinate (Below) at (0,-2.5);
\coordinate (below) at (0,-3.5);
\begin{scope}[very thick,decoration={
    markings,
    mark=at position 0.6 with {\arrow{>}}}
    ] 
\draw [-, dashed,postaction={decorate}] (one) to [bend right=0]  (One);
\draw [-, dashed,postaction={decorate}] (two) to [bend right=0]  (Two);
\draw [-, thick,postaction={decorate}] (One) to [bend right=70]  (Two);
\draw [-, thick,postaction={decorate}] (One) to [bend right=0]  (Two);
\draw [-, thick,postaction={decorate}] (One) to  [bend left=70] (Two);
\draw [-, thick,postaction={decorate}] (Below) to  [bend left=23] (One);
\draw [-, thick,postaction={decorate}] (Two) to  [bend right=-23] (Below);
\draw [-, line width=2pt,postaction={decorate}] (below) to  [bend left=0] (Below);
\end{scope}
\node (d1) at (0,1.05) [font=\scriptsize, text width=.2 cm]{$k_1$};
\node (d2) at (0,0.25) [font=\scriptsize, text width=.2 cm]{$k_2$};
\node (d3) at (0,-0.6) [font=\scriptsize, text width=.2 cm]{$k_3$};
\node (dlp1) at (-1.65,-1.4) [font=\scriptsize, text width=1.8 cm]{$\sum_i k_i-p_1$};
\node (dlp2) at (2.15,-1.4) [font=\scriptsize, text width=1.8 cm]{$\sum_i k_i +p_2$};
\node (p1) at (-2.2,0.25) [font=\scriptsize, text width=1 cm]{$p_1^2=0$};
\node (p2) at (2.4,0.25) [font=\scriptsize, text width=1 cm]{$p_2^2=0$};
\node (p3) at (1.45,-3) [font=\scriptsize, text width=2.4 cm]{$s=(p_1+p_2)^2 $};
\end{tikzpicture}
\caption{The three-loop ice cone graph.}
\label{fig:icecone}
\end{figure}

Our first example of a Feynman graph family with underlying geometry beyond a single elliptic curve is the three-loop 
ice cone family (see figure \ref{fig:icecone}).
%-----Picture Three-Loop Ice Cone Graph-------------
%------------------------

 As it was argued in~\cite{Duhr:2022dxb}, by studying 
the maximal cut of the three-loop ice cone in $d=2$ one finds 
 two  different elliptic curves, which can be both related to the elliptic curve of the two-loop sunrise graph.
We follow the conventions from~\cite{Duhr:2022dxb} and define the propagators and irreducible scalar products
of the ice cone family as
\begin{gleichung}
&\begin{aligned}
     D_1 &= k_1^2-m^2\,,   &D_2 &= k_2^2-m^2\,,  &D_3 &= k_3^2-m^2\,,  \\
     D_4 &= (k_1+k_2+k_3-p_1)^2-m^2\,,  &D_5 &= (k_1+k_2+k_3+p_2)^2-m^2\,,  && \\
     N_1 &= (k_1+k_2+k_3)^2\, , &N_2 &= k_1 \cdot k_3\, , &&
\end{aligned} \\
&N_3 = k_2\cdot k_3\,, \ \   N_4 = k_2 \cdot p_1\,, \ \   N_5 = k_2 \cdot p_2\,, \ \  N_6 = k_3 \cdot p_1\,, \ \  N_7 = k_3 \cdot p_2  \, .
\label{props_ice}
\end{gleichung}
Again following~\cite{Duhr:2022dxb}, we take as the set of starting master integrals\footnote{For simplicity we only write down the first seven $\nu_i$'s. The other numerators are not needed for our choice of master integrals.}
\begin{gleichung}
&\begin{aligned}
    &I_1 = I_{1,1,1,0,0,0,0}\,, \quad &I_2 &=  I_{1,1,1,1,0,0,0}\,, \quad &I_3 &= I_{2,2,0,1,1,0,0}\,, \\
    &I_4 = I_{1,1,1,1,1,0,0}\,, \quad &I_5 &= I_{1,1,1,1,1,-1,0}\,, \quad &I_6 &= I_{2,1,1,1,1,0,0}\,, \quad I_7 = I_{2,1,1,1,1,-1,0}\,,
\end{aligned}\\    
    &I_8 = I_{1,1,1,1,1,-1,-1} +\frac16I_2+\frac16I_4-\frac{1-z+z^2}{6 z}I_5  \, .
\end{gleichung}
To simplify the analysis of the leading singularities and to see the two elliptic curves emerge in $d=2$, we reparametrise the problem through the Landau variable $s=-\frac{(1-z)^2}{z}$ with $s=(p_1+p_2)^2$ and we set $m=1$ for simplicity. 
We then perform a rotation on the integrals $I_4,I_5,I_6,I_7$ to disentangle the two elliptic curves
\begin{gleichung}
    \begin{pmatrix}
        -z & 1 & 0 & 0 \\
        -1 & \frac{1}{z} & -3 & \frac{3}{z} \\
        -\frac{1}{z} & 1 & 0 & 0 \\
        \frac{1}{z^2} & -\frac{1}{z} & \frac{3}{z^2} & -\frac{3}{z}
    \end{pmatrix} \, ,
\end{gleichung}
such that their maximal cuts satisfy a GM system in block form
\begin{gleichung}
\label{GMicehom}
    \textbf{GM}^{\mathcal E_1, \mathcal E_2} = \begin{pmatrix}
        0 & 1 & 0 & 0 \\
        -\frac{1-3 z}{z^2 (1-z) (1-9 z)} & \frac{(1-3 z) (1+3 z)}{z (1-z) (1-9 z)} & 0 & 0 \\
        0 & 0 & 0 & 1 \\
        0 & 0 & \frac{3-z}{z (1-z) (1-9 z)} & -\frac{9-20 z+3 z^2}{z (1-z) (1-9 z)} \\
    \end{pmatrix} \, .
\end{gleichung}
The two blocks correspond to the elliptic Gauss-Manin differential system of the two-loop equal-mass sunrise evaluated at $z$ and $1/z$. For more details, in particular, how this system can be solved in $d=2$, we refer to~\cite{Duhr:2022dxb}.

By simple $\eps$-rescalings of the integrals $I_1,I_2,I_3$ and $I_8$ together with a normalisation by $(1-z^2)/z$ of $I_3$, the corresponding differential equations are brought in $\eps$-factorised form. To achieve an $\eps$-factorised form also for the remaining master integrals $I_4,\hdots,I_7$ we apply our procedure. First, we need the semi-simple part of $\textbf{W}$ corresponding to the Gauss-Manin system~\eqref{GMicehom}. Here we simply take the semi-simple parts of the two equal-mass sunrise Wronskians,
\begin{gleichung}
    \textbf{W}^\text{ss} = \begin{pmatrix}
        \varpi_0 & 0 & 0 & 0 \\
        \varpi_0' & \frac{z}{(1-z) (1-9 z)\varpi_0} & 0 & 0 \\
        0 & 0 & \tilde{\varpi}_0 & 0 \\
        0 & 0 & \tilde{\varpi}_0' & -\frac{9}{z(1-z) (1-9 z)\tilde{\varpi}_0} \\
    \end{pmatrix} \, ,
\end{gleichung}
where 
$$\varpi_0(z) = z(1+3z+15z^2+93z^3)+ \mathcal O(z^4)$$
is the holomorphic solution of the first sunrise block (c.f. eq.\eqref{periodssunrise}) and 
\begin{gleichung}
    \tilde{\varpi}_0(z) = 1+\frac{z}{3}+\frac{5 z^2}{27}+\frac{31 z^3}{243} + \mathcal O(z^4)
\end{gleichung}
is the other holomorphic solution of the second sunrise block. With this ingredient, we construct the rotation matrix to an $\eps$-factorised basis. First, we take the inverse of $\textbf{W}^\text{ss}$ followed by a suitable $\eps$-rescaling. Subsequently, all contributions of $\varpi_0'$ can be removed by another rotation. However, in contrast to the two-loop sunrise graph with equal masses, we need to introduce another rotation to factorise $\eps$ in the part of the differential equations where the two sunrise blocks couple to each other:
\begin{gleichung}
    \begin{pmatrix}
1 & 0 & 0 & 0 \\
 0 & 1 & G & 0 \\
 0 & 0 & 1 & 0 \\
 \frac19G & 0 & 0 & 1
    \end{pmatrix} 
\end{gleichung}
where we introduced a new function defined by
\begin{gleichung}
    G(z)   =   \int_0^z  \left(  \frac{1-18 u+u^2}{u (1+u)^2} \varpi_0(u) \tilde{\varpi}_0(u)  -8 \varpi_0(u)\tilde{\varpi}_0'(u)  \right) \mathrm du \, .
\end{gleichung}
We expect that this function should be related to a new differential form defined on the geometry of the three-loop ice cone graph.
Finally, the non-$\eps$-factorised terms in the contributions from the subsectors can be removed by a final simple rotation. The final $\eps$-factorised differential equation can be found in appendix \ref{app_iceban} eq. \eqref{gmepsice}.

\subsection{The three-loop banana}

It is by now well known that the geometry underlying the $l$-loop banana graph is an $(l-1)$-dimensional \emph{Calabi-Yau variety}~\cite{Kreimer:2022fxm,Fischbach:2018yiu,Klemm:2019dbm,Bourjaily:2018yfy,Bourjaily:2019hmc,Bourjaily:2018ycu,Vergu:2020uur,Duhr:2022pch,Joyce}. It is, therefore, interesting to see if our approach can also help to find $\eps$-factorised bases when such more complicated
geometries are involved.
%-----Picture Three-Loop Banana Graph-------------
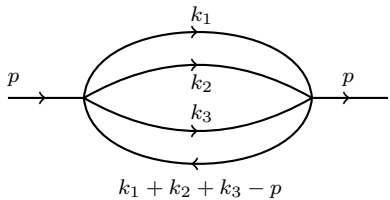
\begin{figure}[!h]
\centering
\begin{tikzpicture}
\coordinate (llinks) at (-2.5,0);
\coordinate (rrechts) at (2.5,0);
\coordinate (links) at (-1.5,0);
\coordinate (rechts) at (1.5,0);
\begin{scope}[very thick,decoration={
    markings,
    mark=at position 0.5 with {\arrow{>}}}
    ] 
\draw [-, thick,postaction={decorate}] (links) to [bend right=30]  (rechts);
\draw [-, thick,postaction={decorate}] (links) to [bend right=-30]  (rechts);

\draw [-, thick,postaction={decorate}] (links) to [bend left=85]  (rechts);
\draw [-, thick,postaction={decorate}] (llinks) to [bend right=0]  (links);
\draw [-, thick,postaction={decorate}] (rechts) to [bend right=0]  (rrechts);
\end{scope}
\begin{scope}[very thick,decoration={
    markings,
    mark=at position 0.5 with {\arrow{<}}}
    ]
\draw [-, thick,postaction={decorate}] (links) to  [bend right=85] (rechts);
\end{scope}
\node (d1) at (0,1.1) [font=\scriptsize, text width=.2 cm]{$k_1$};
\node (d2) at (0,0.2) [font=\scriptsize, text width=.2 cm]{$k_2$};
\node (d3) at (0,-0.2) [font=\scriptsize, text width=.2 cm]{$k_3$};
\node (d4) at (0.45,-1.2) [font=\scriptsize, text width=3 cm]{$k_1+k_2+k_3-p$};
\node (p1) at (-2.0,.25) [font=\scriptsize, text width=1 cm]{$p$};
\node (p2) at (2.4,.25) [font=\scriptsize, text width=1 cm]{$p$};
\end{tikzpicture}
\caption{The three-loop banana graph.}
\label{fig:banana3L}
\end{figure}
%------------------------

As the simplest case, we consider the equal-mass three-loop banana family (see figure \ref{fig:banana3L}) with propagators and numerators:
\begin{gleichung}
\label{props_banL3}
     D_1 &= k_1^2-m^2\,, \   & D_2 &= k_2^2-m^2\,,  & D_3 &= k_3^2-m^2\,,  & D_4 &= (k_1+k_2+k_3-p)^2-m^2\,, \\
     N_1 &= (k_1-p)^2\,, \   & N_2 &= (k_2-p)^2\,,  & N_3 &= (k_3-p)^2\,,  & N_4 &= (k_1-k_2)^2\,, \ N_5 = (k_1-k_3)^2 \, .
\end{gleichung}
A generic integral is then given by eq. \eqref{def:integralfamily}. As customary when dealing with two-point functions, it is convenient
to study the integrals in this family in $d=2-2\eps$ dimensions and as functions of the single dimensionless parameter $z=m^2/p^2$. For the master integrals, we take the dotted basis given by
\begin{gleichung}
\label{masters_ban3L}
    I_1 = I_{1,1,1,0}\,, \quad I_2 = I_{1,1,1,1}\,, \quad I_3 = I_{2,1,1,1} \quad\text{and}\quad I_4 = I_{3,1,1,1} \, .
\end{gleichung}
Notice that to simplify our notation we drop the indices corresponding to the additional numerators
since we do not need them in the following. Moreover, we set $m=1$ to shorten our formulas later.
These master integrals satisfy a non-$\eps$-factorised GM system $\frac{\mathrm d}{\mathrm dz} \underline I = \textbf{GM}\, \underline I\,,$ where the explicit form of $\textbf{GM}$ is given in appendix \ref{app_iceban} eq. \eqref{gmnonban3}. 

The maximal cuts of the three-loop banana master integrals satisfy a GM system which has as solutions the periods of a K3 surface~\cite{Primo:2017ipr,Broedel:2019hyg}. 
As it is well known, differential operators of one-parameter K3 surfaces are 
symmetric squares of elliptic operators~\cite{Doran:1998hm,bogner2013algebraic}.  This was demonstrated explicitly for the differential operator
of the three-loop banana graph in~\cite{Joyce,Primo:2017ipr,Broedel:2019kmn}. The homogeneous $(3\times3)$-system
\begin{gleichung}
    \textbf{GM}^\text{K3} = 
                \begin{pmatrix}
                     \frac{1}{z} & \frac{4}{z} & 0 \\
 -\frac{1}{4 z} & \frac{1-4 z}{4 z^2} & \frac{1}{2 z^2} \\
 \frac{1-28 z+120 z^2}{8 z (1-4 z) (1-16 z)} & -\frac{1-40 z+392 z^2-800 z^3}{8 z^2 (1-4 z) (1-16 z)} & -\frac{1-36 z+264 z^2-256 z^3}{4 z^2 (1-4
   z) (1-16 z)}
                \end{pmatrix}
\end{gleichung}
admits the following solutions
\begin{gleichung}
    \varpi_0(z) &=   z+4 z^2+28 z^3 +256z^4 +\mathcal O(z^5)    \\
    \varpi_1(z) &=   \varpi_0(z) \log(z) + \Sigma_1 = \varpi_0(z) \log(z) + 6 z^2+57 z^3 +584z^4+\mathcal O(z^5) \\
    \varpi_2(z) &=   \frac12\varpi_0(z) \log^2(z) + \Sigma_1\log(z) + 18 z^3 +270z^4+\mathcal O(z^5) \, ,
\end{gleichung}
which can be, if desired, expressed through squares of elliptic integrals as explained in the aforementioned references. For our discussion, this particular form is not needed. Moreover, we hide the explicit dependence of $z$ on the solutions to simplify our formulas. We can construct the Wronskian matrix
\begin{gleichung}
    \textbf{W} = 
        \begin{pmatrix}
             1 & 0 & 0 \\
            -\frac{1}{4} & \frac{z}{4} & 0 \\
            \frac{1}{8} & -\frac{z (1-4 z)}{8}  & \frac{z^3}{2}
        \end{pmatrix}
        \begin{pmatrix}
            \varpi_0 & \varpi_1 & \varpi_2 \\
            \varpi_0' & \varpi_1' & \varpi_2' \\
            \varpi_0'' & \varpi_1'' & \varpi_2''
        \end{pmatrix} \, .
\end{gleichung}
Due to the underlying K3 geometry we get the unipotent part 
\begin{gleichung}
    \textbf{W}^\text{u} =  \begin{pmatrix}
                                         1 & 2\pi i\, \tau  & \frac{(2\pi i)^2}2\, \tau^2\\ 0 & 1 & 2\pi i\, \tau \\ 0 & 0 & 1
                                        \end{pmatrix}  \quad\text{with}\quad   \frac{\mathrm d}{\mathrm d\tau} \textbf{W}^\text{u} = \begin{pmatrix}
                                         0 & 2\pi i & 0\\ 0 & 0 & 2\pi i \\ 0 & 0 & 0
                                        \end{pmatrix} \textbf{W}^\text{u}
\end{gleichung}
and $\tau$-parameter
\begin{gleichung}
  \tau = \frac1{2\pi i} \frac{\varpi_1 }{\varpi_0 } = \frac1{2\pi i}\left( \log(z) + 6 z+33 z^2+284 z^3+\frac{5889 z^4}{2} + \mathcal O(z^5) \right) \, .
\end{gleichung}
From this we can construct the semi-simple part of $\textbf{W}_\text{3L}$ given by
\begin{gleichung}
\label{eq:wrons3L}
     \textbf{W}_\text{3L}^\text{ss} &= 
                \begin{pmatrix}
             1 & 0 & 0 \\
            -\frac{1}{4} & \frac{z}{4} & 0 \\
            \frac{1}{8} & -\frac{z (1-4 z)}{8}  & \frac{z^3}{2}
        \end{pmatrix}  \begin{pmatrix} \varpi_0 & 0 & 0 \\ \varpi_0' & \frac{1}{\sqrt{(1-4 z) (1-16 z)}} & 0 \\ 
                \varpi_0'' & \frac{1}{\sqrt{(1-4 z) (1-16 z)}} \frac{\varpi_0'}{\varpi_0}+\frac{2 (5-32 z)}{((1-4 z) (1-16 z))^{3/2}} & \frac{1}{(1-4 z) (1-16 z)\varpi_0}
                \end{pmatrix} \, .
\end{gleichung}
We can use the Griffith's transversality conditions for the three-loop banana graph~\cite{Bonisch:2021yfw}: 
\begin{gleichung}
\label{eq:griff3L}
    \mathbf Z &= \tilde{\textbf{W}}_\text{3L} \mathbf \Sigma \tilde{\textbf{W}}_\text{3L}^T  \quad\text{with}\quad 
    \tilde{\textbf{W}}_\text{3L} =
    \begin{pmatrix}
            \varpi_0 & \varpi_1 & \varpi_2 \\
            \varpi_0' & \varpi_1' & \varpi_2' \\
            \varpi_0'' & \varpi_1'' & \varpi_2''
        \end{pmatrix}\, ,\quad 
     \mathbf \Sigma = \begin{pmatrix}
         0 & 0 & 1 \\ 0 & -1 & 0 \\ 1 & 0 & 0
     \end{pmatrix} \, , \\
     \mathbf Z^{-1} &=
     \begin{pmatrix}
         \frac{1-8 z}{z^2} & -10+64 z & 1-20 z+64 z^2 \\
 -10+64 z & -1+20 z-64 z^2 & 0 \\
 1-20 z+64 z^2 & 0 & 0
       \end{pmatrix}
\end{gleichung}
to eliminate in eq. \eqref{eq:wrons3L} $\varpi_0''$ in favour of $\varpi_0, \varpi_0$:
\begin{gleichung}
\label{eq:grifrel}
    \varpi_0'' = \frac{1}{2} \left(-\frac{(1-8 z) }{z^2 (1-4 z) (1-16 z)}\varpi_0+\frac{4 (5-32 z) }{(1-4 z) (1-16z)}\varpi_0'+\frac{\varpi_0'^2}{\varpi_0}\right) \, .
\end{gleichung}

Now we can continue as for the equal-mass two-loop sunrise family. We construct our rotation matrix starting with the inverse semi-simple piece of the Wronskian followed by an appropriate $\eps$-rescaling. Notice, that by using  the relation in eq. \eqref{eq:grifrel} we have already removed all contributions of $\varpi_0''$ in the GM equations. We can then proceed to also remove all dependencies of $\varpi_0'$ in the GM system. So far, this is exactly the same procedure as in the two-loop case. Nevertheless, for the three-loop banana, this is not sufficient to achieve full $\eps$-factorisation and  we have to include an extra rotation
\begin{gleichung}
    \begin{pmatrix}
         1 & 0 & 0 & 0 \\
 0 & 1 & 0 & 0 \\
 0 & G_2 & 1 & 0 \\
 0 & -\frac{1}{2}  G_2^2 & - G_2 & 1
    \end{pmatrix}  \begin{pmatrix}
1 & 0 & 0 & 0 \\
 0 & 1 & 0 & 0 \\
 0 & 0 & 1 & 0 \\
 0 & \frac1{\epsilon }G_1 & 0 & 1
    \end{pmatrix} \, ,
\end{gleichung}
where two new functions $G_1$ and $G_2$ have been introduced. They are defined as iterated integrals
\begin{gleichung}
\label{newfunc3L}
     G_1(z)    &=    \int_0^z \frac{2 (1-8 u) (1+8 u)^3}{u^2 (1-4 u)^2 (1-16 u)^2}\, \varpi_0(u)^2\, \mathrm d u    \\
     G_2(z)    &=    \int_0^z \frac{1}{\sqrt{(1-4 u) (1-16 u)}\varpi_0(u)}\, G_1(u)\, \mathrm du
\end{gleichung}
and bring the GM equations into $\eps$-factorised form (see appendix \ref{app_iceban} eq. \eqref{gmepsban3}). This is exactly the same form as found in~\cite{Pogel:2022yat}. Also there the introduction of two new functions given by integrals over the holomorphic solution $\varpi_0$ was necessary. Notice that the new function $G_1$ only appears in the rotation matrix but \emph{not} in the final GM matrix such that only expressions having simple poles appear in the $\eps$-factorised GM matrix as also noticed in~\cite{Pogel:2022yat}.

%% file: Sec5_Conclusions.tex
% !TEX encoding = UTF-8 Unicode

\section{Conclusions and outlook}
\label{sec:conclusions}
In this paper, we have elaborated on a procedure to obtain $\eps$-factorised
systems of differential equations for Feynman integrals which can be expressed
in terms of iterated integrals defined on non-trivial geometries.
We started by summarising physically and mathematically motivated criteria
to select a good basis of master integrals from which to start our procedure.
We have stressed that these criteria do not uniquely constrain
the starting basis, but allow us to choose a set of integrals for which the subsequent
steps of our procedure can be applied more easily.
Once a starting basis has been identified, we proceed bottom-up, sector by sector.
For each sector, the crucial step in our approach consists in
constructing the Wronskian of the corresponding homogeneous differential equations and then
rotating the basis by its semi-simple part. This construction was inspired by findings made
by direct evaluation of elliptic Feynman integrals in terms of pure elliptic multiple polylogarithms~\cite{Broedel:2018qkq}. 
After having rotated by the semi-simple part, one can then proceed by first cleaning
up the homogeneous part of the system and then focusing on the inhomogeneous part. 
These clean-up steps can be seen as a generalisation of the standard approach used for polylogarithmic
integrals and described in detail, for example, in~\cite{Gehrmann:2014bfa}.
The $\eps$-factorised form that we achieve never contains derivatives
of the periods, which we consider an important starting point to prove that the
differential forms are independent, at least on the field of algebraic functions.

An important feature of our approach is that, during the clean-up steps, 
one might have to introduce new non-trivial objects, 
which we refer to as $G_i$ and
cannot simply be expressed as linear combinations of algebraic functions and of the periods of the geometry considered. 
These functions can instead be expressed themselves as \emph{iterated integrals} over periods 
and algebraic functions and they can be related to similar objects
identified in the differential equations for the multi-loop banana graph~\cite{Pogel:2022vat,Pogel:2022ken}.
In our analysis, we have shown that, in many cases, 
the appearance of these new quantities can be related to either
extra master integrals in the homogeneous system for $\eps \neq 0$, or to
extra singular points stemming from subtopologies.

We have successfully applied our procedure to various problems of increasing difficulty, starting from
single-scale elliptic integrals, moving to multi-scale elliptic problems and also to families of integrals which involve geometries beyond an elliptic curve.
In the elliptic case, the interpretation of our results is particularly transparent. As we verified explicitly 
in some of the examples above,
we expect that a new function $G_i$ should be required for each independent 
differential form of the third kind that can be defined on the elliptic curve. 
This can either happen if there are extra master integrals in the top sector, which are linearly dependent in even
integer numbers of dimensions, or if the extra residues are generated by the subtopologies. 
An interesting case in this respect is that of the non-planar three-point functions studied in section~\ref{ssec:triagnle2}. Here, despite the presence of a third master integral in the top topology corresponding to an extra residue, no extra function $G_i$ is required. 
Indeed, we could prove that evaluating that differential form on one of the independent cycles, the integral reduces to a linear combination of
the period and the quasi-period of the curve and it is therefore not linearly independent.

As already discussed, our procedure can also be applied to cases beyond elliptic. 
We considered as examples the three-loop ice-cone graph and the three-loop banana graph. 
Interestingly, also in those cases, the same steps take us to an $\epsilon$-factorised
basis and make it possible to identify in a very clean way a minimal number of new kernels that have to be added to the set of independent differential forms.

While many questions remain to be answered, we believe that our way of constructing
$\epsilon$-factorised bases of master integrals can not only provide a useful tool to solve practical problems but also help shed some light on the general properties of Feynman integrals beyond the polylogarithmic case and their evaluation in terms of iterated integrals
over independent sets of differential forms. As the next steps, we hope to be able to study more in detail the extension of our procedure for either the Calabi-Yau case or higher genera geometries.

%% file: App2_Triangle.tex
\begin{landscape}

\section{Appendix B: Explicit results for triangle graphs}
\label{app_triangle}

Here we give some explicit results for the different triangle families.

The $\eps$-factorised GM matrix of the first triangle family is given by
\begin{gleichung}
\label{gmepstri1}
    \textbf{GM}^{\eps}  =   \left(
\begin{array}{ccccccccccc}
 0 & 0 & 0 & 0 & 0 & 0 & 0 & 0 & 0 & 0 & 0 \\
 0 & \frac{1}{z} & 0 & 0 & 0 & 0 & 0 & 0 & 0 & 0 & 0 \\
 -\frac{1}{z \sqrt{1+4 z}} & 0 & \frac{4 (1+z)}{z (1+4 z)} & \frac{3}{z \sqrt{1+4 z}} & 0 & 0 & 0 & 0 & 0 & 0 & 0 \\
 0 & 0 & -\frac{2}{z \sqrt{1+4 z}} & -\frac{1}{z} & 0 & 0 & 0 & 0 & 0 & 0 & 0 \\
 0 & 0 & 0 & -\frac{1}{z} & 0 & 0 & 0 & 0 & 0 & 0 & 0 \\
 0 & 0 & 0 & 0 & 0 & -\frac{1}{z} & \frac{1}{\sqrt{1-4 z} z} & 0 & 0 & 0 & 0 \\
 \frac{1}{\sqrt{1-4 z} z} & \frac{1}{\sqrt{1-4 z} z} & 0 & 0 & 0 & -\frac{3}{\sqrt{1-4 z} z} & \frac{-3+4 z}{z (-1+4 z)} & 0
   & 0 & 0 & 0 \\
 0 & 0 & 0 & 0 & -\frac{1}{z} & 0 & 0 & 0 & 0 & 0 & 0 \\
 0 & 0 & 0 & 0 & \frac{1}{2 z} & -\frac{1}{z} & 0 & 0 & \frac{1}{z} & 0 & 0 \\
  0 & 0 & 0 & 0 & 0 & 0 & 0 & 0 & 0 & \frac{1-8 z}{z (1-16 z)} & \frac{z^3}{ (1-16 z)\varpi_0(z)^2} \\
 0 & 0 & 0 & \frac{10 \varpi_0(z)}{z^3} & \frac{16 \varpi_0(z)}{z^3} & -\frac{28 \varpi_0(z)}{z^3} & \frac{12
   \varpi_0(z)}{\sqrt{1-4 z} z^3} & 0 & \frac{8 \varpi_0(z)}{z^3} & \frac{ (1-8 z)^2\varpi_0(z)^2}{z^5 (1-16 z)} & \frac{1-8
   z}{z (1-16 z)} \\
\end{array}
\right) \, .
\end{gleichung}

Here we provide the rotation matrix $\textbf{T}^\text{sub}$ which brings the polylogarithmic part of the second triangle family in canonical form
\begin{gleichung}
\label{rottri2}
\textbf{T}^\text{sub}   =    
      \left(
\begin{array}{cccccccccccccccccc}
 -\epsilon ^2 & 0 & 0 & 0 & 0 & 0 & 0 & 0 & 0 & 0 & 0 & 0 & 0 & 0 & 0 & 0 & 0 & 0 \\
 0 & 2 (1-\epsilon ) \epsilon  & 0 & 0 & 0 & 0 & 0 & 0 & 0 & 0 & 0 & 0 & 0 & 0 & 0 & 0 & 0 & 0 \\
 0 & 0 & -s \epsilon ^2 & 0 & 0 & 0 & 0 & 0 & 0 & 0 & 0 & 0 & 0 & 0 & 0 & 0 & 0 & 0 \\
 0 & 0 & 0 & -s \epsilon ^2 & 0 & 0 & 0 & 0 & 0 & 0 & 0 & 0 & 0 & 0 & 0 & 0 & 0 & 0 \\
 0 & 0 & 0 & 2 (1-s) \epsilon ^2 & (1-s) \epsilon ^2 & 0 & 0 & 0 & 0 & 0 & 0 & 0 & 0 & 0 & 0 & 0 & 0 & 0 \\
 0 & 0 & 0 & 0 & 0 & -s \epsilon ^3 & 0 & 0 & 0 & 0 & 0 & 0 & 0 & 0 & 0 & 0 & 0 & 0 \\
 0 & 0 & 0 & 0 & 0 & 0 & -s \epsilon ^3 & 0 & 0 & 0 & 0 & 0 & 0 & 0 & 0 & 0 & 0 & 0 \\
 0 & 0 & 0 & 0 & 0 & 0 & 0 & s \epsilon ^3 & 0 & 0 & 0 & 0 & 0 & 0 & 0 & 0 & 0 & 0 \\
 0 & 0 & 0 & 0 & 0 & 0 & 0 & 0 & -s \epsilon ^3 & 0 & 0 & 0 & 0 & 0 & 0 & 0 & 0 & 0 \\
 \frac{\sqrt{(4+s) s} \epsilon ^3}{2 (1+2 \epsilon )} & 0 & 0 & 0 & 0 & 0 & 0 & 0 & 0 & -s \sqrt{(4+s) s} \epsilon ^2 & 0 &
   0 & 0 & 0 & 0 & 0 & 0 & 0 \\
 0 & 0 & 0 & 0 & 0 & 0 & 0 & 0 & 0 & 0 & -s \epsilon ^4 & 0 & 0 & 0 & 0 & 0 & 0 & 0 \\
 0 & 0 & 0 & 0 & 0 & 0 & 0 & 0 & 0 & 0 & 0 & -s \epsilon ^4 & 0 & 0 & 0 & 0 & 0 & 0 \\
 0 & 0 & 0 & 0 & 0 & 0 & 0 & 0 & 0 & 0 & 0 & 0 & -s \epsilon ^4 & 0 & 0 & 0 & 0 & 0 \\
 0 & 0 & 0 & 0 & 0 & 0 & 0 & 0 & 0 & 0 & 0 & 0 & 0 & -s \epsilon ^3 & 0 & 0 & 0 & 0 \\
 0 & 0 & 0 & 0 & 0 & 0 & 0 & 0 & 0 & 0 & 0 & 0 & 0 & 0 & -s^2 \epsilon ^3 & 0 & 0 & 0 \\
 0 & 0 & 0 & 0 & 0 & 0 & 0 & 0 & 0 & 0 & 0 & 0 & 0 & 0 & 0 & 1 & 0 & 0 \\
 0 & 0 & 0 & 0 & 0 & 0 & 0 & 0 & 0 & 0 & 0 & 0 & 0 & 0 & 0 & 0 & 1 & 0 \\
 0 & 0 & 0 & 0 & 0 & 0 & 0 & 0 & 0 & 0 & 0 & 0 & 0 & 0 & 0 & 0 & 0 & 1 \\
\end{array}
\right) \, .
\end{gleichung}

The $\eps$-factorised GM matrix of the second triangle family is given by
{\begin{tiny}
\begin{gleichung}
\label{gmepstri2}
\textbf{GM}^{\eps}   =    
      \left(
\begin{array}{cccccccccccccccccc}
 0 & 0 & 0 & 0 & 0 & 0 & 0 & 0 & 0 & 0 & 0 & 0 & 0  \\
 0 & 0 & 0 & 0 & 0 & 0 & 0 & 0 & 0 & 0 & 0 & 0 & 0  \\
 0 & 0 & -\frac{1}{z} & 0 & 0 & 0 & 0 & 0 & 0 & 0 & 0 & 0 & 0  \\
 0 & 0 & 0 & \frac{1}{z} & -\frac{1}{-1+z} & 0 & 0 & 0 & 0 & 0 & 0 & 0 & 0  \\
 0 & 0 & 0 & \frac{6}{z} & -\frac{4}{-1+z} & 0 & 0 & 0 & 0 & 0 & 0 & 0 & 0  \\
 \frac{1}{2 (1+z)} & 0 & 0 & \frac{2}{1+z} & -\frac{1}{2 (1+z)} & \frac{1}{z (1+z)} & 0 & 0 & -\frac{1}{1+z} & 0 & 0 & 0 & 0
    \\
 0 & 0 & 0 & 0 & 0 & 0 & \frac{1}{z} & 0 & 0 & -\frac{1}{\sqrt{z (4+z)}} & 0 & 0 & 0  \\
 0 & 0 & 0 & 0 & 0 & 0 & 0 & -\frac{2}{z} & 0 & 0 & 0 & 0 & 0  \\
 -\frac{1}{2 (1+z)} & 0 & 0 & \frac{1-z}{z (1+z)} & \frac{1}{2 (1+z)} & -\frac{1}{z (1+z)} & 0 & 0 & -\frac{2+z}{z (1+z)} &
   0 & 0 & 0 & 0  \\
 -\frac{1}{\sqrt{z (4+z)}} & 0 & -\frac{1}{\sqrt{z (4+z)}} & 0 & 0 & 0 & \frac{3}{\sqrt{z (4+z)}} & 0 & 0 & -\frac{4+3 z}{z
   (4+z)} & 0 & 0 & 0  \\
 0 & 0 & 0 & -\frac{1}{2 z} & 0 & 0 & 0 & 0 & 0 & 0 & 0 & 0 & 0  \\
 0 & 0 & 0 & 0 & 0 & -\frac{1}{z} & \frac{2}{z} & 0 & 0 & 0 & 0 & 0 & 0 0 \\
 0 & \frac{1}{8 z} & 0 & -\frac{3}{2 z} & \frac{1}{4 z} & 0 & 0 & 0 & 0 & 0 & 0 & 0 & \frac{1}{z} 0 \\
 \frac{1}{(1-z) (1+z)} & \frac{1}{4 (1-z)} & 0 & -\frac{1-z+2 z^2}{(1-z) z (1+z)} & \frac{z}{(1-z) (1+z)} & -\frac{2}{(1-z)
   (1+z)} & 0 & 0 & \frac{2 z}{(1-z) (1+z)} & 0 & -\frac{2}{(1-z) z} & 0 & 0  \\
 0 & -\frac{1}{4 (1-z)} & \frac{1}{1-z} & \frac{1}{1-z} & -\frac{1}{2 (1-z)} & 0 & 0 & \frac{2}{1-z} & 0 & 0 & 0 & 0 &
   -\frac{2}{1-z} \\
 0 & 0 & 0 & 0 & 0 & 0 & 0 & 0 & 0 & 0 & 0 & 0 & 0  \\
 0 & -\frac{1}{12 z} & 0 & \frac{4}{3 z} & -\frac{1}{6 z} & -\frac{4}{3 z} & \frac{4}{3 z} & 0 & 0 & 0 & 0 & 0 & -\frac{2}{3
   z}  \\
 \frac{\varpi_0(z)}{2} & -\frac{(8+13 z)\varpi_0(z)}{48 z} & -\varpi_0(z) & \frac{ (20+73 z)\varpi_0(z)}{12 z} &
   -\frac{ (8+37 z)\varpi_0(z)}{24 z} & -\frac{(8-5 z)\varpi_0(z)}{3 z} & \frac{ (32-59 z)\varpi_0(z)}{12 z} & -2 \varpi_0(z) &
   -2 \varpi_0(z) & \frac{ (8+7 z)\varpi_0(z)}{4 \sqrt{z (4+z)}} & -\frac{2 \varpi_0(z)}{z} & 0 & -\frac{ (8+7 z)\varpi_0(z)}{6
   z}  \\
\end{array}
\right. \\[3ex]
\quad \left.
\begin{array}{cccccccccccccccccc}
 0 & 0 & 0 & 0 & 0 \\
 0 & 0 & 0 & 0 & 0 \\
0 & 0 & 0 & 0 & 0 \\
 0 & 0 & 0 & 0 & 0 \\
 0 & 0 & 0 & 0 & 0 \\
  0 & 0 & 0 & 0 & 0 \\
 0 & 0 & 0 & 0 & 0 \\
 0 & 0 & 0 & 0 & 0 \\
 0 & 0 & 0 & 0 & 0 \\
  0 & 0 & 0 & 0 \\
& \frac{1}{z} & 0 & 0 & 0 & 0 \\
 0 & 0 & 0 & 0 & 0 \\
 0 & -\frac{1}{z} & 0 & 0
   & 0 \\
 \frac{3-z}{(1-z) z} & 0 & 0 & 0 & 0 \\
 0 & \frac{2}{1-z} & 0 & 0 & 0 \\
 0 & 0 & \frac{28-2 z+z^2}{3 z (1-z) (8+z)} & 0 & \frac{8}{\varpi_0(z)^2
   z (1-z) (8+z)} \\
 -\frac{2}{3 z} & \frac{2}{3 z} & \frac{4  (1-z)\varpi_0(z)}{9 z} & \frac{2}{3 z} & 0 \\
  \frac{2  (1+2 z)\varpi_0(z)}{3 z} & \frac{ (8+7 z)\varpi_0(z)}{6 z} & \frac{ (2+z) \left(88-84 z+66 z^2+11
   z^3\right)\varpi_0(z)^2}{24 z (1-z) (8+z)} & -\frac{2  (1-z)\varpi_0(z)}{3 z} & \frac{28-2 z+z^2}{3 z (1-z) (8+z)} \\
\end{array}
\right) \, .
\end{gleichung}
\end{tiny}}

\end{landscape}

%% file: App3_Add.tex
\begin{landscape}

\section{Appendix C: Explicit results for ice cone and banana graphs}
\label{app_iceban}

We give in this appendix some explicit results for the ice cone and banana graphs.

The $\eps$-factorised GM matrix for the three-loop ice cone is given by
\begin{gleichung}
\label{gmepsice}
     \textbf{GM}^{\eps}  =   \left(
\begin{array}{cccccccc}
 0 & 0 & 0 & 0 & 0  \\
 0 & 0 & 0 & 0 & 0 \\
 -\frac{2}{z} & 0 & \frac{1-z}{z (1+z)} & 0 & 0  \\
 0 & 0 & 0 & \frac{3-14 z+3 z^2}{z (1-z) (1+z) (1-9 z)} & \frac{z}{ (1-z) (1-9 z)\varpi_0(z)^2}  \\
 0 & \frac{32 \varpi_0(z)}{(1-z) z} & \frac{3 \varpi_0(z)}{z^2} & -\frac{16 \varpi_0(z) G(z)}{ z(9 z) (1-z)\tilde{\varpi}_0(z)}-\frac{G(z)^2}{ z(9-z) (1-z) \tilde{\varpi}_0(z)^2}-\frac{ \left(27-381 z-344 z^2+646 z^3-5173 z^4-1593
   z^5+162 z^6\right)\varpi_0(z)^2}{z^3 (9-z) (1-z) (1+z)^2 (1-9 z)} & \frac{3-14 z+3 z^2}{(1-z) (1+z) (1-9 z)}  \\
 0 & 0 & 0 & -\frac{8 \varpi_0(z)}{ z(9-z) (1-z) \tilde{\varpi}_0(z)}-\frac{G(z)}{ z(9-z) (1-z) \tilde{\varpi}_0(z)^2} & 0  \\
 0 & -\frac{32 \tilde{\varpi}_0(z)}{9 (1-z)} & -\frac{\tilde{\varpi}_0(z)}{3} & \frac{ \left(3-14 z+3 z^2\right)G(z)}{z (9-z) (1-9
   z)}+\frac{2   \left(-9+71 z-35 z^2+5 z^3\right)\varpi_0(z)\tilde{\varpi}_0(z)}{3 z (9-z) (1-z) (1+z)} & \frac{ zG(z)}{9 (1-z) (1-9 z)\varpi_0(z)^2} \\
 0 & 0 & 0 & -\frac{\varpi_0(z)}{6 z^2} & 0  \\
\end{array}
\right. \\
\left.
\begin{array}{cccccccc}
 0 & 0 & 0 \\
 0 & 0 & 0 \\
 0 & 0 & 0 \\
 -\frac{
   zG(z)}{ (1-z) (1-9 z)\varpi_0(z)^2} & 0 & 0 \\
 -\frac{9 
   \left(3-14 z+3 z^2\right)G(z)}{z (9-z) (1-9 z)}-\frac{6 \left(-9+71 z-35 z^2+5 z^3\right)\varpi_0(z)G(z)}{z (9-z)
   (1-z) (1+z)} & \frac{72 \varpi_0(z)}{ z (9-z) (1-z) \tilde{\varpi}_0(z)}+\frac{9 G(z)}{ z (9-z) (1-z)\tilde{\varpi}_0(z)^2} &
   -\frac{72 \varpi_0(z)}{z^2} \\
 -\frac{3-14
   z+3 z^2}{(9-z) (1-z) (1+z)} & \frac{9}{ z (9-z) (1-z)\tilde{\varpi}_0(z)} & 0 \\
 -\frac{ zG(z)^2}{9  (1-z) (1-9 z)\varpi_0(z)^2}-\frac{ \left(-162-441
   z+116 z^2-50 z^3-42 z^4+3 z^5\right)\tilde{\varpi}_0(z)^2}{9 (9-z) (1-z) (1+z)^2} & -\frac{3-14 z+3 z^2}{(9-z) (1-z) (1+z)} & -8 \tilde{\varpi}_0(z) \\
 \frac{\tilde{\varpi}_0(z)}{6} & 0 & -\frac{1+z}{z (1-z)} \\
\end{array}
\right) \, .
\end{gleichung}

The non-$\eps$-factorised GM matrix for the three-loop banana is given by
\begin{gleichung}
\label{gmnonban3}
     \textbf{GM}  =    \left(
\begin{array}{cccc}
 0 & 0 & 0  \\
 0 & -\frac{1+3 \epsilon }{t} & -\frac{4}{t}  \\
 0 & \frac{(1+3 \epsilon ) (1+4 \epsilon )}{4 t} & \frac{4-t+(16-t) \epsilon }{4 t}  \\
 \frac{12 \epsilon ^3}{t (16-t) (4-t)} & -\frac{(1+3 \epsilon ) (1+4 \epsilon ) (8 (15+22 \epsilon )+t (-28+t+(-36+t)
   \epsilon ))}{8 t (16-t) (4-t)} & \frac{t^3 (1+\epsilon )^2-8 t^2 (1+\epsilon ) (5+8 \epsilon )-32 (25+\epsilon  (101+94
   \epsilon ))+8 t (49+\epsilon  (161+124 \epsilon ))}{8 t (16-t) (4-t)}  \\
\end{array}
\right. \\[3ex]
 \quad\left.
\begin{array}{cccc}
 0 \\
 0 \\
 -\frac{1}{2} \\
 -\frac{256 (1+\epsilon )-t (264+384 \epsilon -t
   (36-t+(48-t) \epsilon ))}{4 t (16-t) (4-t)} \\
\end{array}
\right) \, ,
\end{gleichung}
whereas the $\eps$-factorised GM matrix is given by
\begin{gleichung}
\label{gmepsban3}
     \textbf{GM}^{\eps}  &=    \left(
\begin{array}{cccc}
 0 & 0   \\
 0 & \frac{2 (1-10 z)}{z (1-4 z) (1-16 z)}-\frac{G_2(z)}{ \sqrt{(1-4 z) (1-16 z)}\varpi_0(z)}   \\
 0 & -\frac{3 G_2(z)^2}{2  \sqrt{(1-4 z) (1-16 z)}\varpi_0(z)}+\frac{(1+8 z)^2 \left(1-8 z+64 z^2\right)\varpi_0(z)}{2
   z^2 ((1-4 z) (1-16 z))^{3/2}}   \\
 -\frac{24 \varpi_0(z)}{z^2} & \frac{8  (1-8 z) (1+8 z)^3\varpi_0(z)^2}{z^2 (1-4 z)^2 (1-16
   z)^2}+\frac{G_2(z)^3}{ \sqrt{(1-4 z) (1-16 z)}\varpi_0(z)}-\frac{  (1+8 z)^2 \left(1-8 z+64
   z^2\right)\varpi_0(z)G_2(z)}{z^2 ((1-4 z) (1-16 z))^{3/2}}   \\
\end{array}
\right. \\[3ex]
 &\quad\left.
\begin{array}{cccc}
 0 & 0 \\
 \frac{1}{\sqrt{(1-4 z) (1-16 z)}\varpi_0(z)} & 0\\
 \frac{2 (1-10 z)}{z (1-4 z) (1-16 z)}+\frac{2 G_2(z)}{ \sqrt{(1-4 z) (1-16
   z)}\varpi_0(z)} & \frac{1}{\sqrt{(1-4 z) (1-16 z)}\varpi_0(z)}  \\
-\frac{3 G_2(z)^2}{2  \sqrt{(1-4 z) (1-16 z)}\varpi_0(z)}+\frac{
   (1+8 z)^2 \left(1-8 z+64 z^2\right)\varpi_0(z)}{2 z^2 ((1-4 z) (1-16 z))^{3/2}} & \frac{2 (1-10 z)}{z (1-4 z) (1-16
   z)}-\frac{G_2(z)}{ \sqrt{(1-4 z) (1-16 z)}\varpi_0(z)} \\
\end{array}
\right) \, .
\end{gleichung}

\end{landscape}